\documentclass[a4,letters,usenatbib]{mnras}
\usepackage{amsmath}
\usepackage{amssymb} 
\usepackage{multirow}
\usepackage{textcomp}
\usepackage{microtype}
\usepackage{subcaption}
\usepackage[dvipsnames]{xcolor}
\usepackage{ulem}
\usepackage{graphicx}
\usepackage{float}
\usepackage{mathptmx}
\usepackage{tabularx}
\usepackage{caption}

\newcommand\numberthis{\addtocounter{equation}{1}\tag{\theequation}}

\newcommand{\email}[1]{\mbox{\href{mailto:#1}{#1}}}

\newcommand{\Mc}{\mathcal{M}}
\newcommand{\Msol}{M_{\odot}}

\newcommand{\Mstar}{{\cal M}_{*}}
\newcommand{\Adet}{A_{\mathrm{det}}}
\newcommand{\Aul}{A_{\mathrm{ul}}}
\newcommand{\Atrial}{A_{\mathrm{trial}}}
\newcommand{\ndot}{{\dot n}_0}

\newcommand{\sigul}{\sigma_{\mathrm{ul}}}
\newcommand{\sigdet}{\sigma_{\mathrm{det}}}

\newcommand{\nunit}{\mathrm{Mpc}^{-3} \mathrm{Gyr}^{-1}}

\def\ltsima{$\; \buildrel < \over \sim \;$}
\def\simlt{\lower.5ex\hbox{\ltsima}}
\def\gtsima{$\; \buildrel > \over \sim \;$}
\def\simgt{\lower.5ex\hbox{\gtsima}}
\def\apj{Astrophysical Journal}                
\def\apjl{Astrophysical Journal}             

\def\apjs{ApJS}              
\def\mnras{MNRAS}            
\def\prd{Phys.~Rev.~D}       
\def\araa{ARA\&A}             
\def\nat{Nature}              
\def\aap{A\&A}                

\title[Multiscale history of massive BH binaries]{Probing the assembly history and dynamical evolution of massive black hole binaries with pulsar timing arrays}
\author[Chen et al.]{
Siyuan Chen,$^1$\thanks{E-mail: \email{schen@star.sr.bham.ac.uk}}
Hannah Middleton,$^1$\thanks{E-mail: \email{hannahm@star.sr.bham.ac.uk}}
Alberto Sesana,$^1$
Walter Del Pozzo$^{1,2}$
\newauthor and Alberto Vecchio$^1$
\\
$^1$School of Physics \& Astronomy, University of Birmingham, Birmingham, B15 2TT, UK\\
$^2$Dipartimento di Fisica ``Enrico Fermi'', Universit\`a di Pisa, Pisa I-56127, Italy
}

\date{Accepted \dots Received \dots; in original form \dots}

\pubyear{2017}

\begin{document}
\label{firstpage}
\pagerange{\pageref{firstpage}--\pageref{lastpage}}
\maketitle

\begin{abstract}
We consider the inverse problem in pulsar timing array (PTA) analysis, investigating what astrophysical information about the underlying massive black hole binary (MBHB) population can be recovered from the detection of a stochastic gravitational wave background (GWB). We employ a physically motivated model that connects the GWB spectrum to a series of parameters describing the underlying redshift evolution of the MBHB mass function and to the typical eccentricity they acquire while interacting with the dense environment of post merger galactic nuclei. This allows the folding in of information about the spectral shape of the GWB into the analysis. The priors on the model parameters are assumed to be uninformative and consistent with the current lack of secure observations of sub-parsec MBHBs. We explore the implications of current upper limits as well as of future detections with a variety of PTA configurations. We confirm our previous finding that current upper limits can only place an upper bound on the overall MBHB merger rate. Depending on the properties of the array, future detections can also constrain several MBHB population models at different degrees of fidelity. In particular, a simultaneous detection of a steepening of the spectrum at high frequency and a bending at low frequency will place strong constraints on both the MBHB mass function and on the typical eccentricity of inspiralling MBHBs, providing insights on MBHB astrophysics unlikely to be achievable by any other means.  
\end{abstract}


\section{Introduction}
\label{sec:Introduction}

Massive black holes (MBHs) appear to be a fundamental component in galaxy formation and evolution. In fact, all massive galaxies appear to host MBHs in their centres \citep[][and references therein]{2013ARA&A..51..511K}. In the hierarchical clustering model of structure formation \citep{1978MNRAS.183..341W}, these MBHs are the dormant counterparts of quasars and active galactic nuclei \citep[e.g.][]{2006ApJS..163....1H}. In a nutshell, galaxies grow through a sequence of mergers and accretion episodes that trigger star formation and fuel the central MBHs. Gas accretion powers luminous electromagnetic radiation, which is at the basis of the Quasar phenomenon \citep[e.g.][]{2006MNRAS.365...11C}. If most galaxies host MBHs then, following galaxy mergers, the two MBHs sink to the center of the merger remnant eventually forming a bound MBH binary \citep[MBHB][]{1980Natur.287..307B}. The details of this general picture are not well understood. In particular, it is not clear whether MBHBs efficiently merge as a consequence of galaxy mergers and what the details of the dynamical processes driving their final coalescence are \citep[see][and references therein]{2012AdAst2012E...3D}.

MBHBs are among the loudest sources of gravitational waves (GWs) in the Universe, and during their inspiral emit radiation that falls in the nHz frequency range, probed by ongoing and upcoming pulsar timing array (PTA) experiments \citep{SesanaVecchioColacino:2008}. In fact, GWs imprint a distinctive signature in the time of arrivals (ToAs) of ultra-stable millisecond pulsars (MSPs). This signature can be disentangled from other noise sources by cross-correlating ToA time-series from an ensemble of pulsars \citep{HellingsDowns:1983}. PTAs therefore monitor a large number of MSPs, looking for this distinctive correlation \citep{1990ApJ...361..300F}. This challenge is currently undertaken by the European Pulsar Timing Array \citep[EPTA][]{2016MNRAS.458.3341D}, the Parkes Pulsar Timing Array \citep[PPTA][]{2016MNRAS.455.1751R} and North American Nanohertz Observatory for Gravitational Waves \citep[NANOGrav][]{2015ApJ...813...65T}. The three collaborations are joining forces under the aegis of the International Pulsar Timing Array \citep[IPTA][]{2016MNRAS.458.1267V}, paving the way towards a future global collaboration that will take advantage of upcoming facilities such as the South African telescope array MeerKAT \citep{2009arXiv0910.2935B}, the Chinese 500-mt telescope FAST \citep{2011IJMPD..20..989N} and eventually the Square Kilometre Array (SKA).

Since PTAs observe individual pulsars with cadence $\Delta t$ of order of few weeks for an experiment duration $T$ of several years, they are most sensitive to GWs in the frequency range $1/\Delta t<f<T$, i.e. $10^{-9}-10^{-7}$ Hz. At such low frequencies, the superposition of GW signals emitted by a cosmological population of MBHBs results in a stochastic GW background (GWB), although, especially at high frequencies, particularly massive/nearby systems may be resolved individually \citep{SesanaVecchioVolonteri:2009}. The GWB amplitude and spectral shape depends on the underlying population of MBHBs and can therefore be used to constrain their astrophysical and cosmological properties \citep{Sesana:2013CQGra}. 

Direct detection of GWs by Advanced LIGO recently opened the high frequency GW window on the Universe \citep{2016PhRvL.116f1102A,2016PhRvL.116x1103A}. Besides the profound implications for gravity theory and fundamental physics \citep{2016PhRvL.116v1101A}, from an astrophysical perspective, GWs are a new tool to understand the physics of compact objects populating the Universe, and how they connect with the evolution of gas, stars and galaxies. From this point of view, PTAs provide a formidable tool to understand the dynamics of MBHBs and their demographic along the cosmic history. In fact, the amplitude of the GWB depends on how frequently MBHBs merge and what their typical mass is, whereas the spectral shape also depends on the mechanism driving the MBHB inspiral and, crucially, on their eccentricity. It is well known that under the assumption of circular GW driven binaries, the characteristic GWB strains follows a power-law $h_c\propto f^{-2/3}$ \citep{Phinney:2001}. However, at large orbital separations (i.e. at low frequencies), MBHB evolution is dominated by energy and angular momentum exchange with the stellar and gas rich surroundings, potentially growing the MBHB eccentricity and resulting in a low frequency turnover of the GWB \citep{2007PThPh.117..241E,2011MNRAS.411.1467K,Sesana:2013CQGra,2014MNRAS.442...56R,2016MNRAS.tmp.1488K,2016arXiv160607484R}. Therefore, the characterization of the amplitude and spectral shape of the GWB carries precious information on the underlying population of MBHBs. To what extent such information can be recovered via PTA observations is the main focus of this paper.

PTA's effort has been so far focusing on delivering the best possible ToA datasets \citep[e.g.][]{2016MNRAS.458.1267V} and on developing the necessary data analysis tools for detection of either a GWB or individual sources \citep[e.g.][]{2012ApJ...756..175E,2013PhRvD..87f4036P,2013PhRvD..87j4021L}. The application of the latter to the former resulted so far in upper limits only \citep{2015MNRAS.453.2576L,2016MNRAS.455.1665B,2016ApJ...821...13A}, and in the absence of a detection, little effort has been spent in the `inverse problem', namely on investigating what astrophysical information can be recovered from PTA observations. This does not mean that astrophysics has been so far ignored; for example, \cite{2016ApJ...821...13A} discussed in length the consequences of their upper limit for MBHB dynamics, and \cite{2016ApJ...826...11S} explored the implications for MBH mass-galaxy relations proposed in the literature. However, although astrophysical inference has been applied to specific upper limits, a framework that connects PTA observations to MBHB astrophysics in the general context of any PTA detection is missing. As part of the common effort of the EPTA collaboration \citep{2016MNRAS.458.3341D} to detect GWs with pulsar timing, this paper is an attempt of making a step forward towards the creation of such a framework. \cite{2016arXiv161202817T} provides an independent, parallel and complementary investigation using Gaussian process emulation techniques.

We consider the model developed in \cite{2016arXiv161200455C}, hereinafter PaperI, for the GWB emitted by a generic population of eccentric MBHBs evolving via scattering of ambient stars. In our model, MBHBs hold a constant eccentricity so long as their evolution is driven by stellar scattering, and circularize under the effect of GW radiation when their dynamics is GW driven (i.e., after decoupling from the stellar environment). In PaperI we showed that the decoupling radius is only a mild fraction of the density of ambient stars, and for stellar density typical of massive galaxies, occurs at frequencies well below the relevant PTA range. As such, we found that the effect of eccentricity is much more prominent, therefore the GWB shape can be fully characterized by a few parameters defining the mass function of MBHBs and its redshift evolution, and the typical eccentricity at decoupling. Expanding on  \cite{2016MNRAS.455L..72M} (M16 hereinafter), we simulate GWB detection for a variety of PTAs and we investigate to what extent the underlying MBHB population parameters can be constrained.

The paper is organized as follows. In section \ref{sec:model} we summarize the relevant features of the GWB spectral models developed in PaperI. In section \ref{sec:simspectrum} we introduce the theory of GWB detection with PTAs and define the impact of the relevant array quantities on the signal-to-noise ratio (S/N) of the measurement. The set-up of our simulations is outlined in section \ref{sec:simobs} and the analysis method used for astrophysical inference is described in section \ref{sec:DAmethod}. We present and discuss in detail our results in section \ref{sec:Results} and conclude with some final remarks and prospects for future expansion of this work in section \ref{sec:Conclusions}.

\section{Astrophysical model}
\label{sec:model}

We use the model developed in PaperI for a population of eccentric MBHBs evolving via three-body scattering against the stellar environment. In PaperI, we expressed the properties of the environment (stellar density, velocity dispersion etc.) as a function of the MBHB total mass only; therefore, the MBHB mass defines the relevant stellar background properties, which we take to be consistent with that typical of elliptical galaxies (where the most massive binaries, dominating the GWB, reside). In a nutshell, the stellar density is modelled with a Dehnen profile \citep{1993MNRAS.265..250D} with total mass set by the intrinsic relation between the MBH and the galaxy bulge masses -- usually referred to as $M_{\rm BH}-M_{\rm bulge}$-- provided in \cite{2013ARA&A..51..511K}, scale radius $a$ defined by the empirical $M_{\rm bulge}-a$ relation found by \cite{2008MNRAS.386..864D}{\footnote{We note that this relation connects the scale radius $a$ to the total mass of the system. However, the massive elliptical galaxies that host the dominant PTA GW sources, are bulge dominated so that $M_{\rm bulge}$ can be taken as a fair proxy of the total stellar mass.}}, and inner profile slope $\gamma=1$, appropriate for massive ellipticals. In this model, binaries decouple from the stellar environment at orbital frequencies much lower than the relevant PTA window (which is $f>1$nHz) and the PTA signal can be constructed taking into account the post-decoupling GW-driven evolution of the eccentric binary only (see PaperI for a full description of the model). The overall GWB spectrum can therefore be written as: 
\begin{equation}
\begin{split}
  h_c^2(f) = & \int dz \int d{\cal M} \frac{d^2n}{dzd{\cal M}} h_{c,{\rm fit}}^2\Big(f\frac{f_{p,0}}{f_{p,t}}\Big) \\ & \times \Big(\frac{f_{p,t}}{f_{p,0}}\Big)^{-4/3} \Big(\frac{\mathcal{M}}{\mathcal{M}_0}\Big)^{5/3} \Big(\frac{1+z}{1+z_0}\Big)^{-1/3}
  \label{eqn:hsquared}
\end{split}
\end{equation}
where $h_{c,{\rm fit}}$ is an analytic fit to the spectrum produced by a reference binary at redshift $z_0$ with chirp mass $\Mc_0$ and a given eccentricity $e_0$ at an arbitrary decoupling frequency $f_0$. These two latter parameters define the peak frequency of the emitted GW spectrum $f_{p,0}$ for this reference binary. Equation (\ref{eqn:hsquared}) states that the overall GW spectrum from a given MBHB population can be generated from this reference $h_{c,{\rm fit}}$ via appropriate power-law scaling of the the chirp mass, redshift, decoupling frequency and eccentricity. Individual contributions must then be integrated over the MBHB mass function $d^2n/dzd\Mc$; the number of binary mergers per co-moving volume, redshift and (rest-frame) chirp mass interval. The integration limits of equation (\ref{eqn:hsquared}) are set to $0\leq z\leq 5$ and $10^6\leq\Mc/\Msol\leq10^{11}$, and following M16 we pick
\begin{align*}
\frac{d^2n}{dzd\log_{10}\Mc} = {\dot n}_0\, \left[\left( \frac{\Mc}{10^7\Msol}\right)^{-\alpha} \exp^{-(\Mc/\Mstar)}\right]\\ \times 
\left[(1+z)^{\beta} \exp^{-(z/z_*)}\right] \frac{dt_R}{dz}\,, \numberthis 
\label{eqn:modeldNdVdzdlogM}
\end{align*}
where $t_R$ is the time in the source rest-frame and $dt_R/dz$ is given by the standard time-redshift cosmological relation (in this work we assume $H_0 =70\mathrm{km~s}^{-1}\mathrm{Mpc}^{-1}$, $\Omega_M=0.3$, $\Omega_{\Lambda}=0.7$ and $\Omega_k=0$). The differential merger rate density of equation (\ref{eqn:modeldNdVdzdlogM}) is described by five parameters. ${\dot n}_0$ is the merger rate density normalization. $\beta$ and $z_*$ describe the redshift evolution of the rate. In particular, $\beta$ controls the low-redshift power-law slope and $z_*$ the high-redshift cut-off for the distribution; the peak of the merger rate corresponds to a redshift ($z_* \beta -1$).  $\alpha$ and $\Mstar$ are the free parameters of the Schechter function describing the mass distribution. In addition to those, the computation of the GWB in equation (\ref{eqn:hsquared}) requires the specification of the MBHB eccentricity $e_t$ when they decouple from their environment and the evolution is dominated by GW emission\footnote{In this pilot study, we make the simplistic assumption that all MBHBs have the same eccentricity at decoupling. In general, MBHBs are expected to have a range of eccentricities when they decouple from their environment. Nonetheless, one can still try to model the population with a single parameter $e_t$, representing the typical MBHB eccentricity.}, giving a total of six model parameters. Decoupling takes place when the condition that stellar scattering and GW emission extract energy from the MBHB at the same rate. This occurs at a frequency $f_t$, defined by (see PaperI)
\begin{equation}
  f_t = 0.356\, {\rm nHz}\, \left(\frac{1}{F(e)}\frac{\rho_{i,100}}{\sigma_{200}}\right)^{3/10}\mathcal{M}_9^{-2/5},
  \label{eq:fdec}
\end{equation}
where the mass density of the stellar environment is $\rho_{i,100}=\rho_i/(100\,\Msol{\rm pc}^{-3})$, the velocity dispersion of the stars is the bulge is $\sigma_{200}=\sigma/(200\,{\rm km\,s}^{-1})$ and the MBHB total mass is $\mathcal{M}_9=\mathcal{M}/(10^9\,\Msol)$. Expressions for $\rho_{i,100}$ and $\sigma_{200}$ can be found in PaperI (equations 28 and 30). Note that $\rho_{i}$ is a function of the inner slope of the adopted density profile. Here we adopt a Dehnen model with $\gamma=1$, which results in shallow nuclear stellar density profiles that are typical of massive elliptical galaxies.

The characteristic amplitude described by equation (\ref{eqn:hsquared}) is a power-law with a low frequency turnover due to eccentricity and environmental effects. At high frequency, however, because of small number statistics, the actual signal is characterized by sparse resolvable systems outshining the overall GWB. \cite{SesanaVecchioColacino:2008} showed that the correct estimate of the unresolved GWB level can be recovered by setting an upper limit $\bar\Mc$ to the mass integral given by the condition

\begin{equation}
  N_{\Delta{f}}=\int_{f-\Delta f/2}^{f+\Delta f/2}df \int_{\bar\Mc}^{\infty}d\Mc \int_{0}^{\infty}dz \frac{d^3 N}{df dz d\mathcal{M}} = 1,
  \label{eq:Nf}
\end{equation}
where $d^3 N/(df dz d\mathcal{M})$ is the number of individual sources per unit chirp mass, redshift and frequency, which can be directly computed from $d^2n/dzd\Mc$ \citep[see][for details]{SesanaVecchioColacino:2008}, and the integral is performed over the frequency bin $\Delta{f}=1/T$. The net effect is that the spectrum has a mass function dependent high frequency steepening, that can provide further information about the underlying MBHB population. Note that this is set solely by the MBHB mass function and does not introduce further parameters to the model. Examples of spectra highlighting both the low frequency turnover and the high frequency steepening are shown in Fig. \ref{fig_hc}.

The model was chosen to capture the expected qualitative features of the cosmic MBH merger rate without restricting to any particular merger history; for example, it can reproduce rates extracted from merger tree models \citep{2003ApJ...582..559V,SesanaVecchioColacino:2008}, and large scale cosmological simulations of structure formation \citep{SpringelEtAl_MilleniumSim:2005,SesanaVecchioVolonteri:2009}.

\section{Background detection theory}
\label{sec:simspectrum}
The S/N $\rho$ imprinted by stochastic GWB in a PTA can be written as \citep{2015CQGra..32e5004M,2015MNRAS.451.2417R}
\begin{equation}
\rho^2=2\sum_{i=1,N}\sum_{j>i}T_{ij}\int \frac{\Gamma_{ij}^2S_h^2}{(S_n^2)_{ij}}df.
\label{eqrho}
\end{equation}
We now proceed to define and discuss all the elements appearing in equation (\ref{eqrho}). $T_{ij}$ is the time span over which observations for pulsars $i$ and $j$ overlap. We will make from here on the simplifying assumptions that all pulsars are observed for the same timespan $T$ (typically 10 years or more) and therefore $T_{ij}=T,\, \forall (i,j)$. However, we should bear in mind that this is generally not the case for real PTAs. The double sum runs over all the possible pairs of pulsars in the array and $\Gamma_{ij}$ are the Hellings \& Downs correlation coefficients \citep{HellingsDowns:1983}
\begin{equation}
\label{eq:gamma}
\Gamma_{ij}=\frac{3}{2}\gamma_{ij}\ln \left(\gamma_{ij}\right)-\frac{1}{4}\gamma_{ij}+\frac{1}{2}+\frac{1}{2}\delta_{ij},
\end{equation}
where $\gamma_{ij}=[1-\cos(\theta_{ij})]/2$, and $\theta_{ij}$ is the relative angle between pulsars $i$ and $j$. $S_h,S_n$ are the spectral densities of the signal and the noise respectively. The former is connected to the characteristic amplitude of the signal $h_c(f)$ given in equation (\ref{eqn:hsquared}) via:
\begin{equation}
\label{eq:sh}
S_h=\frac{h_c^2}{12\pi^2f^3},
\end{equation}
where $f$ is the considered frequency. The latter has to be handled with care, especially in the limit of a strong GWB signal. For a pulsar $i$ characterized by random Gaussian irregularities described by a root mean square (rms) value $\sigma_i^2$, the power spectral density (PSD) of the noise is given by
\begin{equation}
\label{eq:pisimple}
P_i=2\sigma_i^2\Delta t,
\end{equation}
where $\Delta t$ is the interval between subsequent observations (typically a week to a month, in current PTAs). If red processes were not present in the data, one might then expect a PSD of the noise equal to $P_i$ in the whole sensitivity window down to $1/T$. However, fitting for the spin first and second derivatives when constructing the pulsar timing model subtracts a quadratic function to the timing residual, effectively absorbing power at the lowest frequency bins, should a red signal be present.

To mimic the effect of the timing model we empirically write 
\begin{equation}
\label{eqpicorr}
P_i=2\sigma_i^2\Delta t+\frac{\delta}{f^5},
\end{equation}
where $\delta$ is a constant that depends on the parameters of the observations. We find that a good fit to the low frequency behaviour of the published EPTA, NANOGrav and PPTA sensitivity curves is provided by setting
\begin{equation}
\label{eqpicorr}
\delta=5\times10^{-49} \left(\frac{10{\rm yr}}{T}\right)^5\left(\frac{\sigma_i}{100{\rm ns}}\right)^2\frac{\Delta t}{2 {\rm weeks}}.
\end{equation}
The scaling in equation (\ref{eqpicorr}) ensures that the curve maintains the same shape when varying the array parameters, reproducing the power absorption at the two lowest frequency bins (see Fig. \ref{fig_hc}). Moreover the PSD of the noise $S_n$ is not only given by limitations in the pulsar stability, quadratic spindown fitting, and other sources of noise. The very same signal $S_h$ contributes an equal amount to the noise as to the signal itself, because half of the GWB (the pulsar term) is uncorrelated. However, the smoking-gun of a GWB is provided by its distinctive quadrupole correlation described by the $\Gamma_{ij}$ coefficients. Therefore only the correlated part of the signal (i.e. the Earth term) contributes to the construction of the detection statistic and to the build-up of the S/N. The pulsar term will just produce an uncorrelated common red noise in all pulsars with PSD $S_h$. Therefore the power spectral density of the noise has to be written as \citep{2015MNRAS.451.2417R}:
\begin{equation}
  S_{n,ij}^2=P_iP_j+S_h[P_i+P_j]+S_h^2(1+\Gamma_{ij})^2.
    \label{eqsn}
\end{equation}
Note that equation (\ref{eqsn}) reduces to $S^2_{n,ij}=P_iP_j$ in the weak signal limit. Note, moreover, that this implies that it does not matter how strong the signal is, the integrand of equation (\ref{eqrho}) is at most of the order $\Gamma_{ij}^2\ll 1$. This means that {\it only} with a large number $N$ of pulsars is it possible to produce a confident detection of a GWB with an high $\rho$. This is easy to see if we make the simplifying assumptions that $T$, $\Delta t$ and $\sigma_i$ are the same for all pulsars. Moreover, we shall assume a sufficiently high number of randomly distributed pulsars in the sky, therefore substituting the individual $\Gamma_{ij}$ with their average value $\Gamma=1/(4\sqrt{3})$. Equation (\ref{eqrho}) can then be written as
\begin{equation}
\rho^2=2T\Gamma^2 \int \frac{S_h^2}{S_n^2} \sum_{i=1,N}\sum_{j>i} df,
\end{equation}
which reduces to 
\begin{equation}
\rho^2=T\Gamma^2N(N-1)\int \frac{S_h^2}{S_n^2} df.
\label{eqrhosimpl}
\end{equation}
In an actual observation, the GWB is resolved in bins $\Delta f=1/T$. We can therefore divide the frequency domain in intervals $\Delta f_i=[i/T,(i+1)/T]$ centred at $f_i=(2i+1)/(2T)$ and compute the S/N in each individual frequency bin as\begin{equation}
\rho^2_i=T\Gamma^2N(N-1)\int_{\Delta f_{i}} \frac{S_h^2}{S_n^2} df \approx \Gamma^2N(N-1)\frac{S_h^2}{S_n^2}
\label{eqrhobin}
\end{equation}
The total S/N of the observation is then simply obtain by summing in quadrature over the frequency bins
\begin{equation}
\rho=\left(\sum_i \rho_i^2\right)^{1/2}.
\label{eqrhotot}
\end{equation}
Note that in the limit of $S_h\gg P$ in a given frequency bin, equation (\ref{eqrhobin}) reduces to
\begin{equation}
\rho^2_i=\frac{\Gamma^2}{1+\Gamma^2}N(N-1).
\label{eqrhobin}
\end{equation}
Therefore, in the presence of a strong signal in $M$ frequency bins, one gets an approximate S/N 
\begin{equation}
\rho=\left(\frac{\Gamma^2}{1+\Gamma^2}MN(N-1)\right)^{1/2}\approx \Gamma N M^{1/2}.
\label{totrhosimple}
\end{equation}
Where we used the fact that $\Gamma\ll1$ and $N\gg1$. Equation (\ref{totrhosimple}) was obtained through a number of drastic simplifications, nonetheless it gives a sense of the maximum S/N one can obtain assuming a strong signal in an ideal array. Since $\Gamma\approx 0.14$, a total S/N$\approx 5$ in the lowest few frequency bins can only be achieved with approximately $N=20$ equally good pulsars. 

\section{Simulating observations}
\label{sec:simobs}
Once $\rho_i$ has been computed at each frequency bin, we can then use the general fact that, if $h$ is a signal described by an amplitude $A$, then $\rho=(h|h)$ and $\sigma_A^{-1}=(\partial{h}/\partial{A},\partial{h}/\partial{A})^{1/2}=(h/A,h/A)^{1/2}$. Therefore
\begin{equation}
\frac{\sigma_{A}}{A}=\sigma_{{\rm ln}A}=\frac{1}{\rho}.
\label{eqrhotot}
\end{equation}
To simulate observations, we therefore compute the S/N $\rho_i$ at each frequency bin. If $\rho_i>1$, we then assume an observed signal with amplitude $A_i=h_c(f_i)$ and error described by a log-normal distribution with width given by equation (\ref{eqrhotot}). Note that, by doing this we are ignoring any stochastic fluctuation in the measured amplitude of the signal. In reality, the error on the observation will be generally centred at $A_i\neq h_c(f_i)$, with a scatter of the order of the error on the measurement. We make this choice because our main aim is to investigate to what level the MBHB population model can be constrained {\it in principle}, independent of statistical variations inherent to the observations. If $\rho_i<1$ then we assume no signal is detected in the frequency bin, and only an upper limit can be placed. To define what the upper limit is, we notice that, by means of equation (\ref{eq:sh}), equation (\ref{eqrhobin}) can be written as a ratio of the characteristic signal and an equivalent characteristic noise, i.e.,
\begin{equation}
\rho_i=\frac{h_c^2}{h_n^2}
\end{equation}
where, 
\begin{equation}
h_n=[N(N-1)]^{1/4}\left(12\pi^2f^3\frac{S_n}{\Gamma}\right)^{1/2}.
\label{eqhn}
\end{equation}
Therefore, when $\rho_i<1$ we place a 68\% (1$\sigma$) upper limit at $h_{n,i}$, calculated at the central frequency $f_i$ of the bin.

\begin{figure}
\centering
\includegraphics[width=8.cm,clip=true,angle=0]{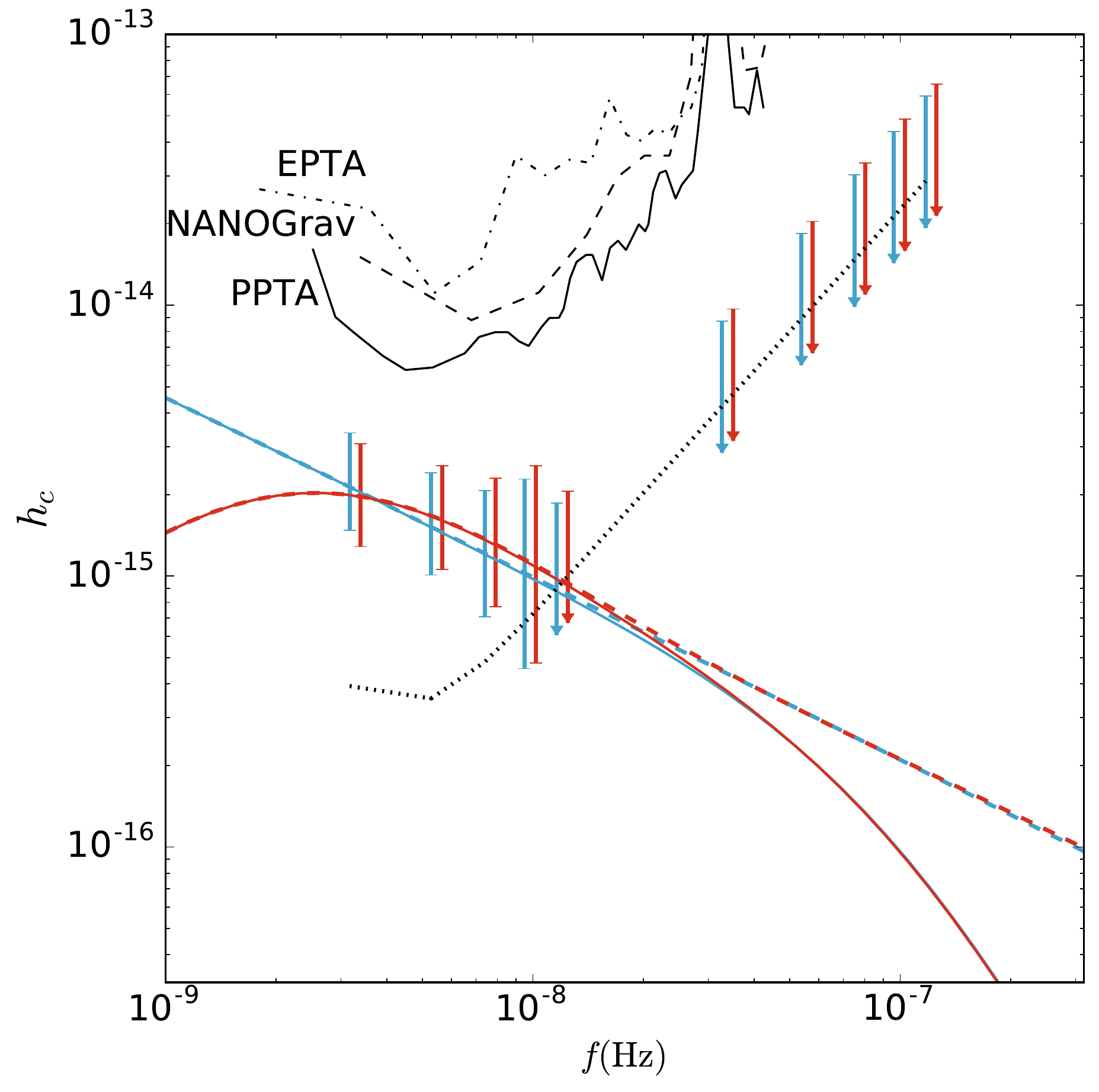}
\caption{Examples of simulated detections for two different spectral shapes. Signal models correspond to the default MBHB population with parameters defined in Section \ref{sec:simsetup} and high eccentricity ($e_t=0.9$, red) and almost circular ($e_t=0.01)$, blue). For each model, solid lines are the theoretical spectra including the high frequency steepening due to the mass upper limit defined by equation (\ref{eq:Nf}), dashed lines depict spectra excluding this feature (therefore with $h_c\propto f^{-2/3}$ at high frequency) for comparison. Error bars centred around the model value are the observed amplitudes with associated uncertainties when $\rho_i>1$, and downward arrows represent upper limits equal to $2h_n$ (i.e. 2$\sigma$) when $\rho_i<1$ at their base. The black dotted line is the characteristic noise level $h_n$ {\it excluding} the contribution of the GW signal to the noise budget. Black lines in the upper part of the figure are current EPTA, NANOGrav and PPTA limits. We assume 15 years of observation of 20 pulsars at 100ns rms.}
\label{fig_hc}
\end{figure}

Examples of signal generation are shown in Fig. \ref{fig_hc} for spectra with 
$A=10^{-15}$ at $f=$1/1yr and an array with $N=20$, $\sigma=100$ns, $T=15$yr, $\Delta t=1$ week. This setup results in a detection with moderate S/N, $\rho\approx 5$, and with $\rho_i\approx 2$ in the few lowest frequency bins. The equivalent $h_n$ of equation (\ref{eqhn}) is depicted as a black solid line. Note, however that for clarity of representation, we ignored here the contribution of $S_h$ to the noise (when that is taken into account, $h_n=h_c$ whenever $\rho_i>1$). Note also that, despite the large $h_c$ difference of the two signals, the difference in S/N between them is only about 20\%. This is because, as stressed above, in the strong signal limit the S/N of the signal is limited by the GWB uncorrelated self-noise.

\subsection{Simulation setup}
\label{sec:simsetup}
To setup a specific simulation, one has to define both the properties of the GWB (i.e. the six parameters ${\dot n}_0, \beta, z_*, \alpha, \Mstar, e_t$ defining the MBHB population) and of the PTA employed for detection (i.e. the four parameters $N, \sigma, T, \Delta t$ defining the sensitivity of the array). 

Unless otherwise stated, we use a MBHB mass function defined by ${\dot n}_0=10^{-4} {\rm Mpc}^{-3}{\rm Gyr}^{-1}, \beta=2, z_*=2, \alpha=0, \Mstar=10^8\Msol$. The normalization ${\dot n}_0$ and the redshift dependence $\beta$ are chosen to be consistent with current estimates of the galaxy merger rate \citep{2004ApJ...617L...9L,2009A&A...498..379D,2011ApJ...742..103L}. $\alpha$ and $\Mstar$ are chosen to ensure that the shape of the MBHB mass function is consistent with that of nuclear MBHs as inferred from direct measurements and MBH-galaxy scaling relations \citep[e.g.][]{2004MNRAS.354.1020S,2007ApJ...654..731H}. The adopted parameters result in a GWB with characteristic strain at $f=1$yr$^{-1}$ of $A\approx5\times10^{-16}$, fully consistent with current upper limits. We explore different eccentricities at decoupling and we report results for the illustrative cases of quasi circular and highly eccentric binaries, defined by $e_t=0.01$ and $e_t=0.9$ respectively.

We make the simplifying assumption that all pulsars are observed for the same timespan $T$, with the same cadence $\Delta t$ and have the same rms $\sigma$. Note that our main results are nevertheless general, since these assumptions only affect the computation of the S/N and do not enter in the subsequent analysis of the GWB spectral shape. We consider four different array scenarios:
\begin{enumerate}
\item case {\it PPTA15}: in this case we simply use the curve provided by \citep{2015Sci...349.1522S}, which is representative of current PTA capabilities and results in an upper limit of $A=10^{-15}$. 
\item case {\it IPTA30}: $N=20$, $\sigma=100$ns, $T=30$yr, $\Delta t=1$ week. This PTA results in a detection S/N$\approx 5$ and is based on a future extrapolation of the current IPTA, without the addition of new telescopes. 
\item case {\it SKA20}: $N=100$, $\sigma=50$ns, $T=20$yr, $\Delta t=1$ week. This PTA results in a high significance detection with S/N$\approx 30-40$, which will be technically possible in the SKA era.
\item case {\it ideal}: $N=500$, $\sigma<1$ns, $T=30$yr, $\Delta t=1$ week. This is likely beyond SKA capabilities but provides useful insights of what might be achievable in principle.
\end{enumerate}  

As stated above, for each simulations we compute the the S/N $\rho_i$ at each frequency bin. If $\rho_i>1$, we then assume an observed signal with amplitude $A_i=h_c(f_i)$ and error described by a log-normal distribution with width given by equation (\ref{eqrhotot}). If $\rho_i<1$ then we place an upper limit at $h_n$ as defined by equation (\ref{eqhn}).

\section{Data analysis method}
\label{sec:DAmethod}

As in M16, our aim is to constrain the astrophysical population of merging MBHB given some PTA data. The data consists of an array of measurements and upper limits on the GW spectrum at different frequency bins, as described in the previous section. In M16, we assumed circular binaries and an $f^{-2/3}$ power law for the spectrum, meaning that all the information from the background could be summarised with two numbers, an upper limit or detection with some confidence at a given frequency, which we chose to be one over one year. In this paper, we allow for eccentric binaries evolving via scattering of background stars and a finite number of sources at high frequencies, both of which result in a spectrum that is different from the $f^{-2/3}$ power law. Therefore, the shape of the spectrum over the frequency band encodes much more information. In this section, we describe our strategy to infer the astrophysical properties of the merging MBHB population from PTA measurements. 

We denote our astrophysical model (section \ref{sec:model}) as $M$ and our data (section \ref{sec:simsetup}) as $d$. Our intention is to infer the model parameters $\theta$, given a specific measurement. We start from Bayes theorem,
\begin{equation}
 p(\theta|d,M) = \frac{p(\theta| M) p(d|\theta,M)}{p(d|M)}\,,
 \label{eqn:bayes}
\end{equation}
where $p(\theta|d,M)$ is the posterior distribution for the model parameters given the data and the model, $p(\theta|M)$ is the prior, representing any initial knowledge we have on the parameters given the specific model, $p(d|\theta,M)$ is the likelihood for the data given the model and some values of the parameters and finally $p(d|M)$ is the evidence.

As described in section \ref{sec:model}, our model has six parameters $\theta={\ndot,\beta,z_*,\alpha,\Mstar,e_t}$. Unless otherwise stated, for our analysis we choose priors as follows: the parameters $\beta$, $z_*$, $\alpha$, $\log_{10}\Mstar$ and $e_t$ are all uniformly distributed in the ranges $\beta \in [-2,7]$, $z_* \in [0.2,5]$, $\alpha \in [-3,3]$, $\log_{10}\Mstar/\Msol \in [6,11]$, and $e_t \in [10^{-6}, 0.999]$. The prior for the merger rate parameter, $\ndot$ is log-uniform for $\ndot \in [10^{-20},10^{3}]$ and uniform in $\ndot$ for $\ndot < 0$, thus allowing for the possibility of no mergers. We note that although specific combinations of parameters can mimic MBHB merger rates extracted from semi-analytic merger tree models \citep{SesanaVecchioColacino:2008}, cosmological simulations of galaxy formation \citep{SesanaVecchioVolonteri:2009,2016MNRAS.tmp.1488K} and observations of galaxy pairs \citep{Sesana:2013}, the adopted prior range is highly uninformative and allows for exotic MBHBs mass functions that are not necessarily related to galaxy mergers. For example, the upper limit in $\ndot$ is solely dictated by the constraint that all the dark matter in the Universe is formed by merging MBHs.

The functional form of the likelihood function we adopt depends upon the type of data in each frequency bin. For a given spectrum, there are two possible observational outcomes in a specific frequency bin $f$; either a GWB detection at $\Adet(f)$, or a non-detection, resulting in an upper limit based on the PTA sensitivity at that frequency $\Aul(f)$. In the case of an upper limit $\Aul(f)$ on the GWB, we model the likelihood as a smooth step-like distribution which allows $\Atrial(f)\ll\Aul(f)$ and tails off to $0$ for $\Atrial(f)\gg\Aul(f)$. For that we use a Fermi-like distribution,
\begin{equation}
 p_\mathrm{ul}(d|\Atrial(f)) \propto \left\{ \exp\left[\frac{\Atrial(f)-\Aul(f)}{\sigul(f)}\right]+1\right \} ^{-1},
 \label{eq:likeul}
\end{equation}
where $\Atrial(f)$ is the GWB given by our model for a set of parameters drawn from the prior and $\sigma_{ul}(f)$ controls the width and steepness of the distribution as it transits at the step $\Aul(f)$ from some constant value for $\Atrial(f)\ll\Aul(f)$ to $0$ at $\Atrial(f)\gg\Aul(f)$. $\sigma_{ul}(f)$ can be adjusted so that, for example $p(\Atrial(f)<\Aul(f))=68\%$. In our simulations, $\Aul(f)=h_n$ as described in section \ref{sec:simobs}. We are therefore using the sensitivity of the PTA as a proxy for the 68\% (or \textit{1-sigma}) upper limit when the signal is not detected.

In the case of a GWB detection of a central amplitude $\Adet(f)$ with a Gaussian distribution width of $\sigdet(f)$, we apply a Gaussian in the logarithm for the likelihood,
\begin{equation}
p_\mathrm{det}\left(d|\Atrial(f) \right) \propto \exp\left\{- \frac{\left[ \log_{10}\Atrial(f)-\log_{10}\Adet(f)\right]^2}{2\sigdet(f)^2} \right\} \,,
\label{eq:likedet}
\end{equation}
where $\sigdet(f)$ is the error on the detection measurement as described in section \ref{sec:simobs} and $\Atrial(f)$ is again the value of the GWB given by parameters sampled by the prior. As the dataset $d$ consists of a collection of GWB measurements across the frequency spectrum, we need to combine the likelihood of all the frequency bins in our data. We assume statistical independence among the various frequency bins and thus compute the overall likelihood by multiplication of the likelihoods (either an upper limit or a detection) from each bin. Note that, when we combine bins with detections to bins with upper limits, we consider the lowest frequency upper limit and five further points spaced by ten bins. This is because bins become much denser at high frequency and considering all the upper limits slows done the likelihood computation substantially. We checked that this does not affect our results, since the only constraining upper limit is always the one at the lowest frequency.

We explore the parameter space by means of a Nested Sampling algorithm~\citep{Skilling2004a}. We use a tailored version of the parallel implementation of Nested Sampling given in~\cite{cpnest} which is similar in spirit to~\cite{VeitchVecchio:2010} and \cite{LALInference}. For all the analysis presented in this work we set the number of live points to be $N \sim$ 2,000 owing an average number of posterior samples $\sim$ 5,000.

\section{Results and discussion}
\label{sec:Results}

In this section we present and discuss in detail the results of our simulations. We will start with the interpretation of upper limits and then move to the case of detections with small and large S/N. We stress that, unless otherwise stated, astrophysical interpretation is constructed uniquely on the basis of PTA observations, i.e. we do not use any additional constraints on the MBHB population (besides the wide, non-informative prior range of the model parameters). PTA inference can prove significantly more constraining if combined with independent information. For example one can assume a narrow prior on the MBHB merger rate and mass function based on simulations or observations of merging galaxies \citep[e.g.][]{Sesana:2013}. However, we caution that such information is often {\it indirect} and requires theoretical modelling subject to several assumptions.  

\subsection{Upper limits}

\begin{figure*}
\includegraphics[width=0.4\textwidth]{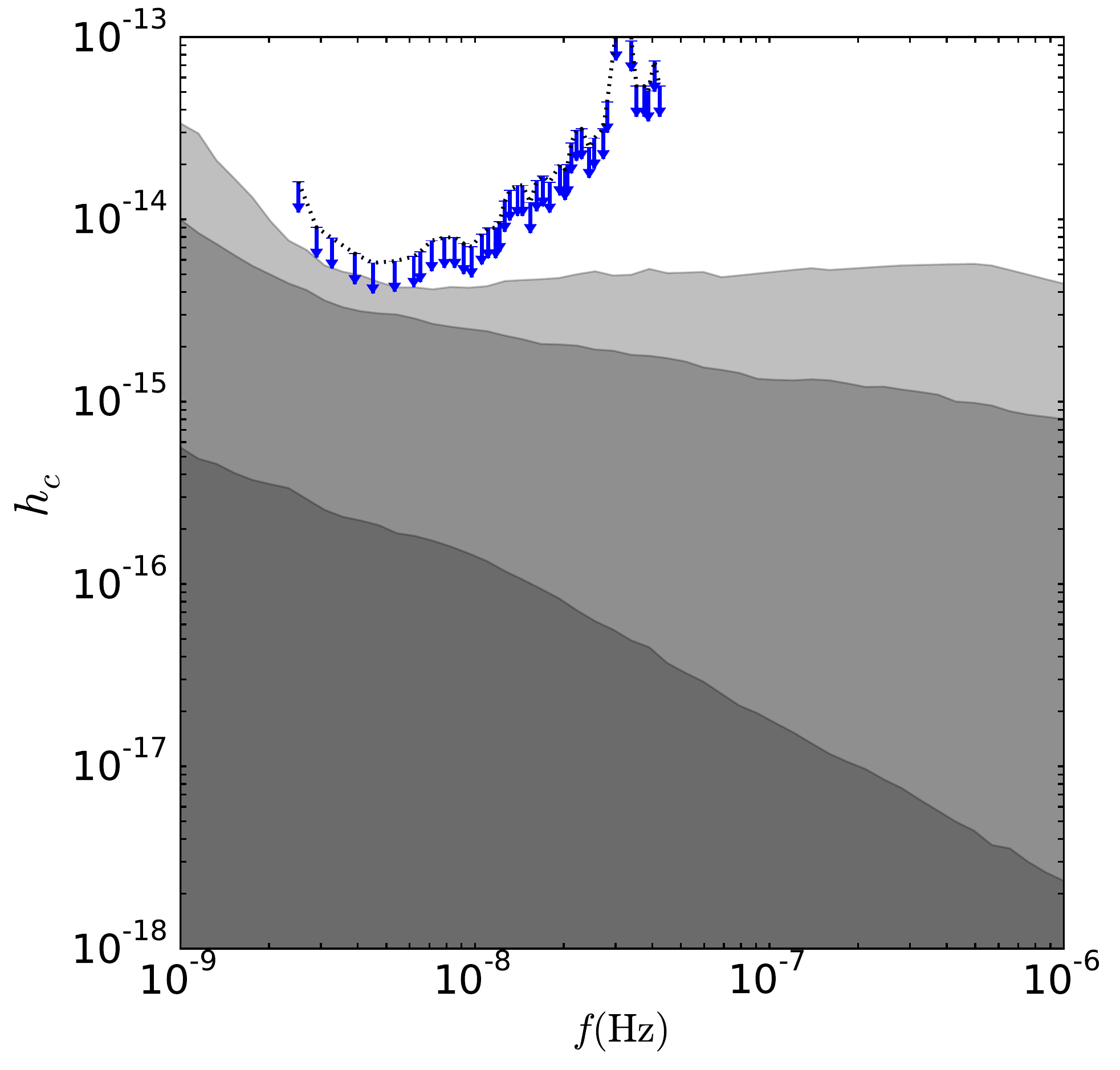} \hspace{1cm}
\includegraphics[width=0.4\textwidth]{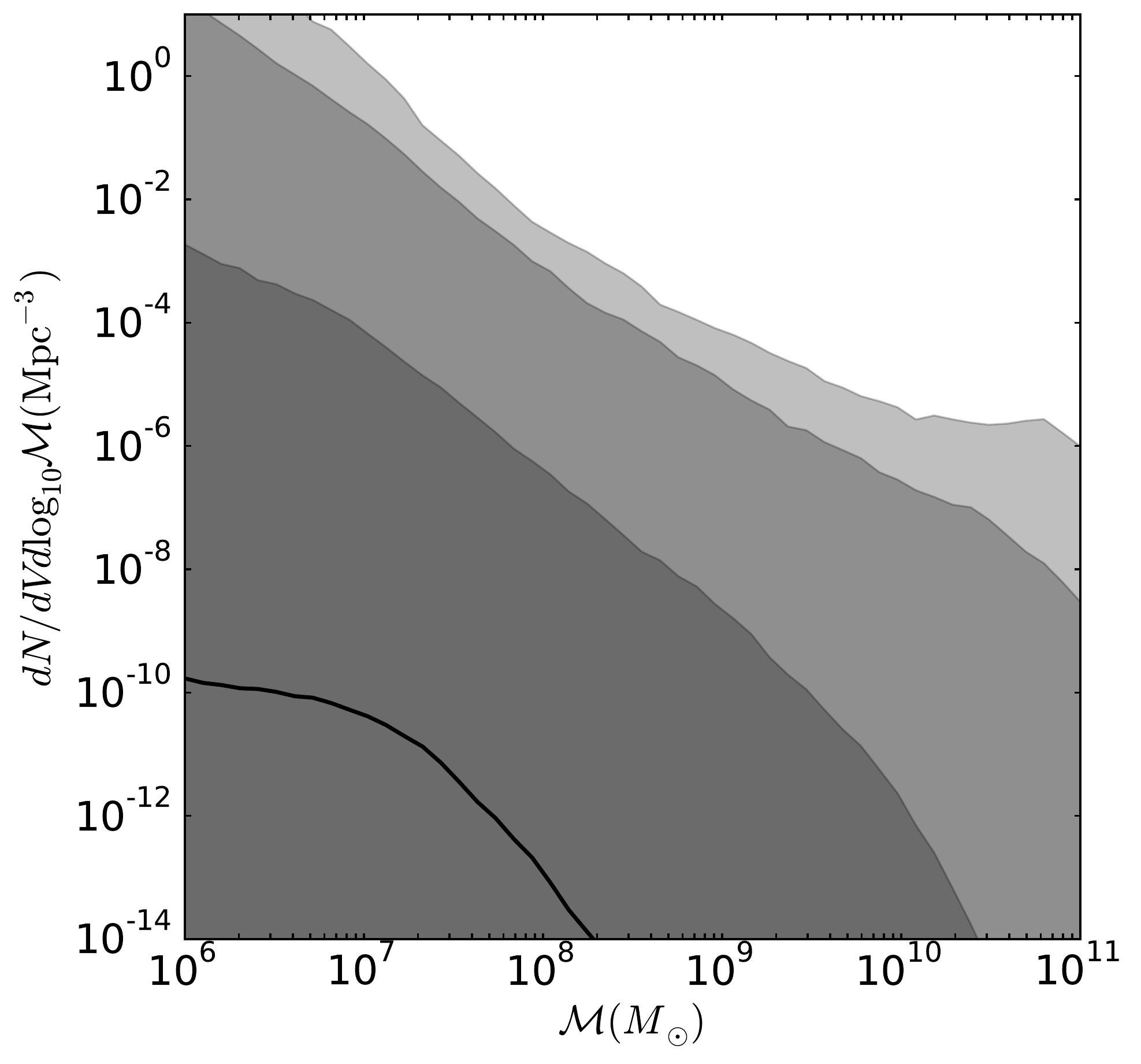}\\
\includegraphics[width=0.4\textwidth]{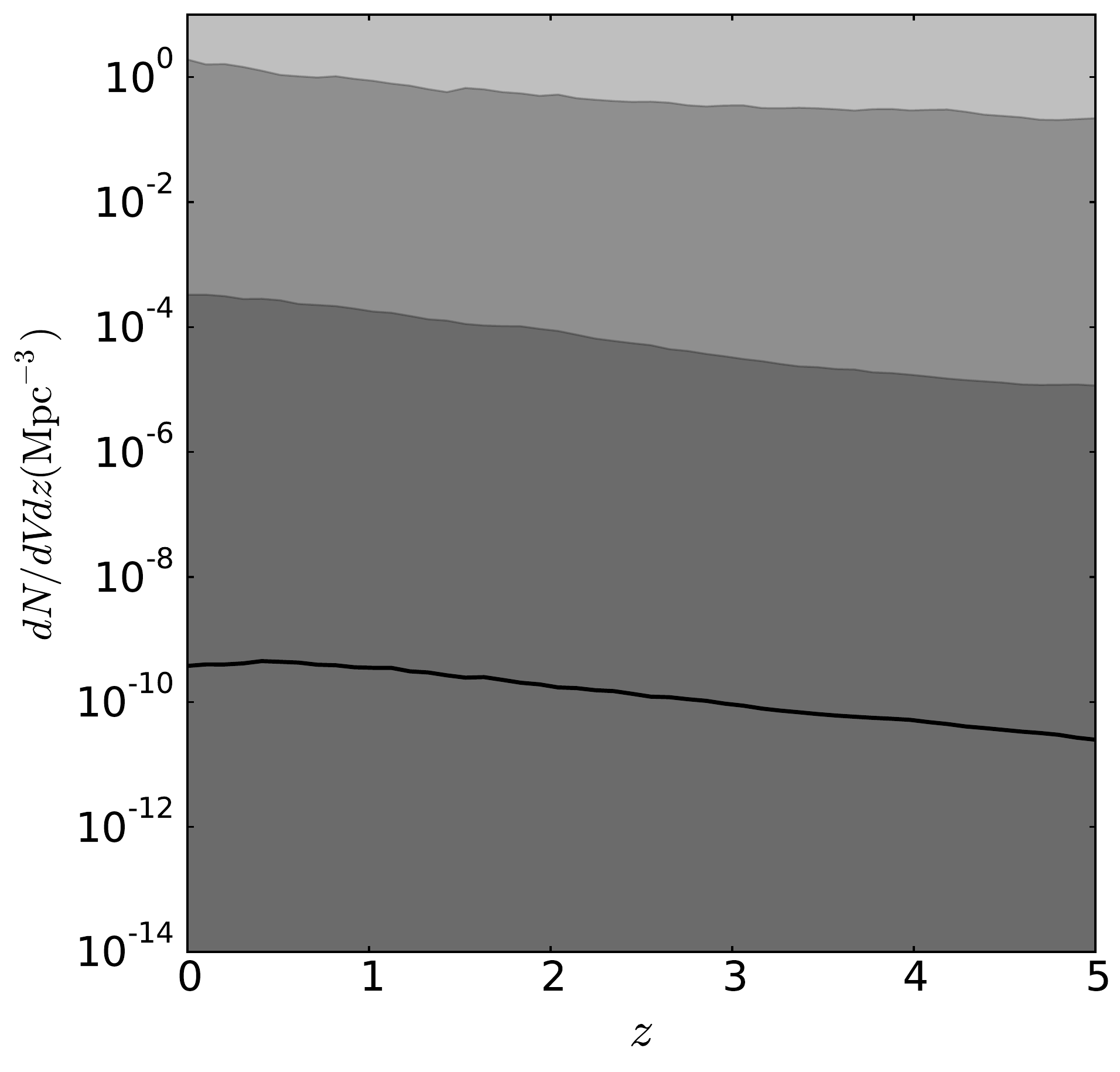} \hspace{1cm}
\includegraphics[width=0.4\textwidth]{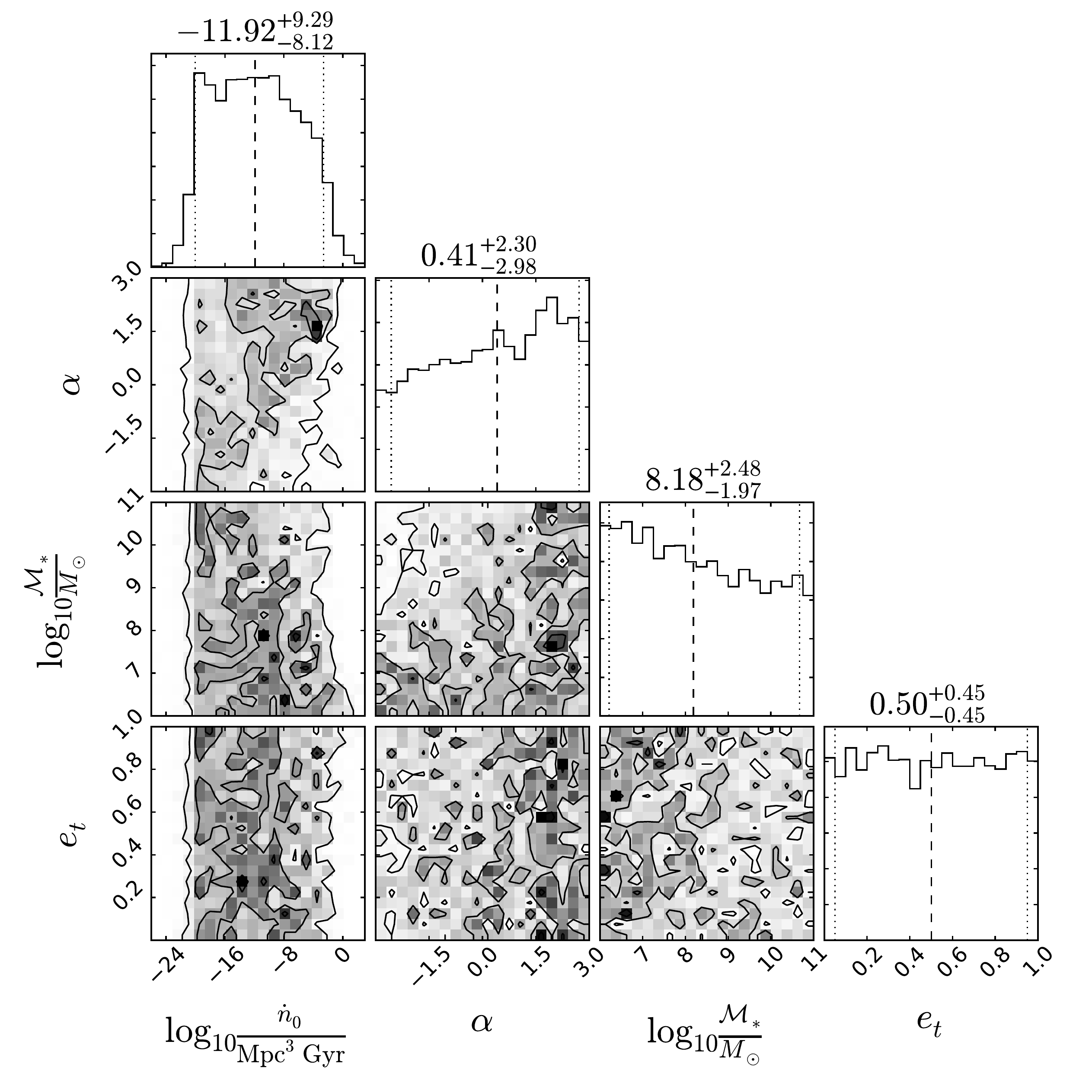}\\
\caption{Implication of a 95\% upper-limit of $A(f=\mathrm{yr}^{-1})= 1\times 10^{-15}$, which corresponds to the most stringent PTA upper limit to date. The posterior for the spectrum (top left), mass (top right) and redshift functions (bottom left) are shown as shaded areas, with the 68\%, 95\% and 99.7\% confidence regions indicated by progressively lighter shades of grey, and the solid black line marking the median of the posterior. The dotted line with downwards pointing arrows in the top left panel is the 95\% upper limit from \protect\cite{2015Sci...349.1522S}. The bottom right triangular plot shows the two-dimensional posteriors for each model parameter pairs, together with their one dimensional marginalised distributions. The lines in each one dimensional distribution mark the median (dashed) and the central 90\% (dotted) of the posterior, with the numerical values indicated above each plot.}
\label{fig:upperlimit}
\end{figure*}

We first consider the case of an upper limit and we take as example the most stringent constraint imposed by the PPTA of $A<10^{-15}$ at $f=1$yr$^{-1}$. Although PTAs often quote limits at $f=1$yr$^{-1}$, those are the result of the integrated array sensitivity across the relevant frequency band. This is shown in the upper-left panel of Fig. \ref{fig:upperlimit}; according to the analysis framework developed in section \ref{sec:model}, we assume at each frequency bin a 95\% upper-limit given by the dashed curve and run our analysis. Consistent with M16, the results shown in Fig. \ref{fig:upperlimit} indicate that current PTA upper limits alone return little astrophysical information, and only loose upper bounds can be placed on the MBHB mass function (upper-right panel) and redshift (lower-left panel) distribution. Those are defined by integrating equation (\ref{eqn:modeldNdVdzdlogM}) in the redshift range [0,5] and in the mass range [$10^6\Msol$,$10^{11}\Msol$], respectively. The triangle plot in the lower-left panel shows that the posterior distributions of the model parameters are essentially flat ($\beta$ and $z_*$ are not shown, as they are always flat due to strong degeneracy with $\ndot$), with the exception of $\ndot$, which is found to be $<2.5\times10^{-3}\nunit$ at the 95\% level. This constraint becomes interesting when compared to independent information on galaxy merger rates. Several observational studies place the merger rate density of massive galaxies at $z<1$ to be around few$\times 10^{-4}$Mpc$^{-3}$Gyr$^{-1}$ \citep{2004ApJ...617L...9L,2011ApJ...742..103L,2012ApJ...747...85X}. In fact, this is in essence the reason why some tension between PTA upper limits and vanilla MBHB assembly models was highlighted by \cite{2015Sci...349.1522S}. We will return in more depth on this point in a companion paper (Middleton et al. in preparation). A tighter upper limit, constraining $\ndot$ to be less than $10^{-5}\nunit$ might rule out a naive one-to-one correspondence between galaxy and MBHB mergers, indicating that delays, stalling or high MBHB eccentricities play a major role in the dynamics.

\subsection{PTA detection constraints on model parameters}

\begin{figure*}
\begin{subfigure}{0.425\textwidth}
\centering
$IPTA30, \ e_t = 0.01$
\includegraphics[width=6.0cm]{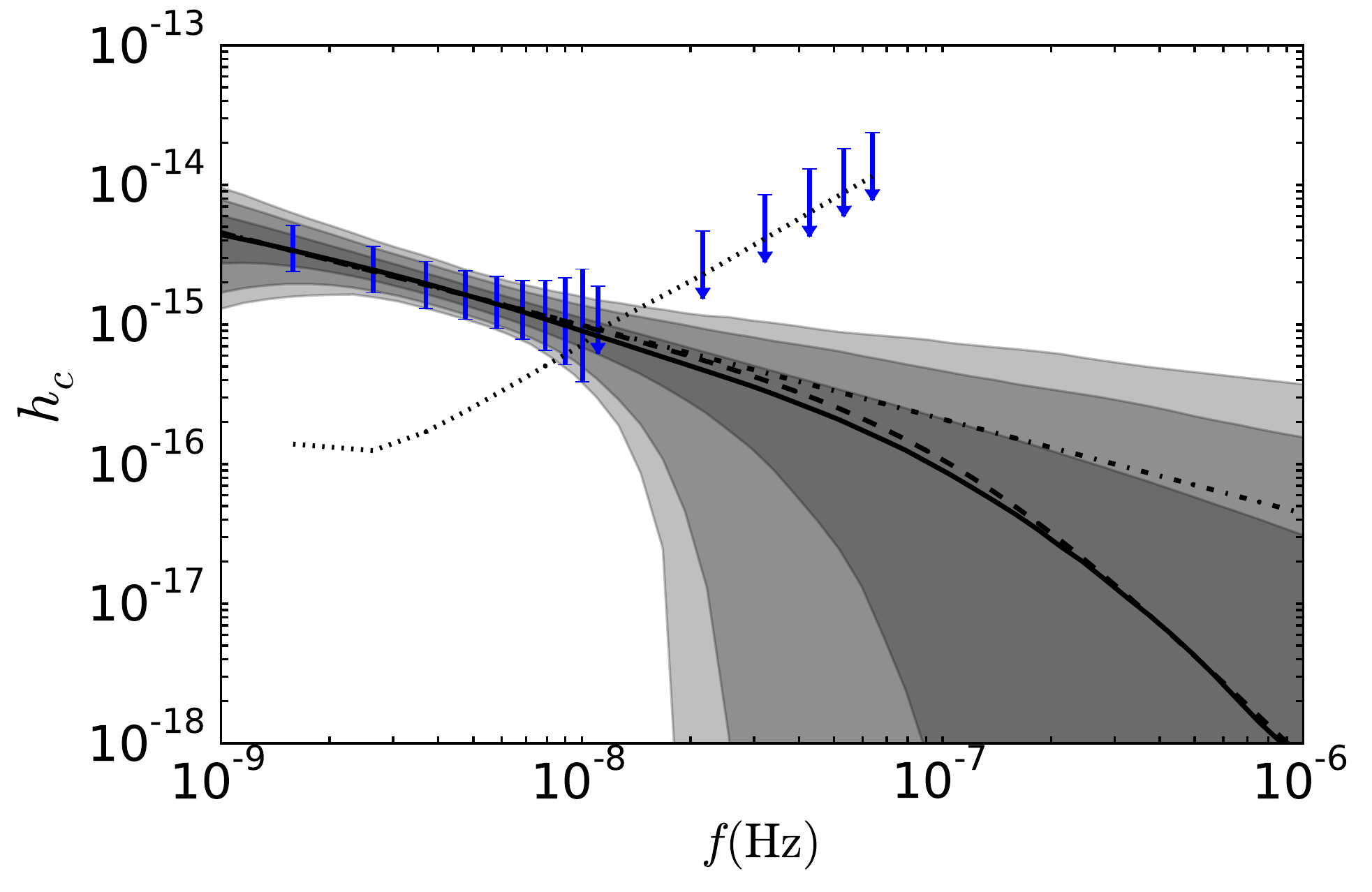}\\
\includegraphics[width=6.0cm]{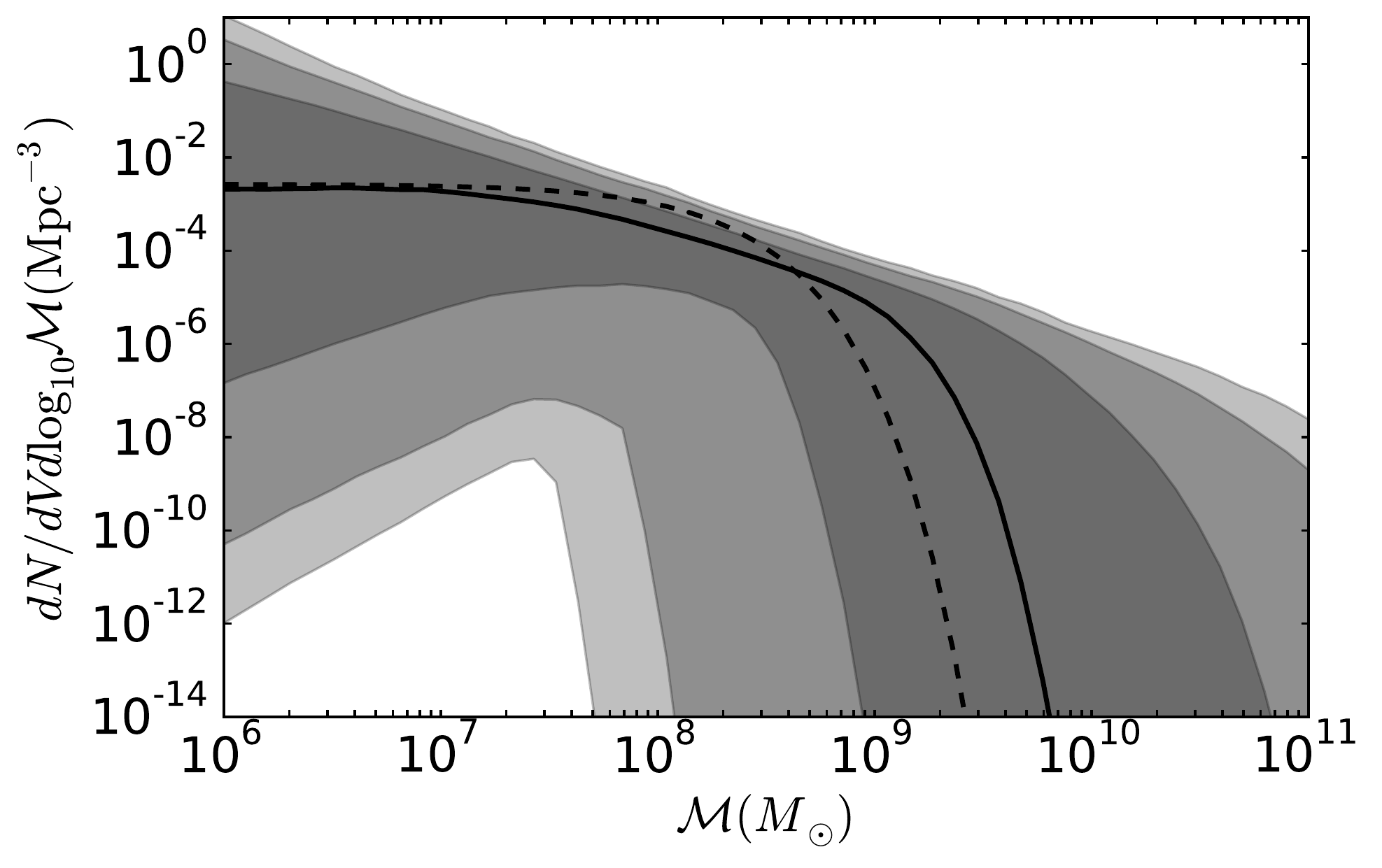}\\
\includegraphics[width=6.0cm]{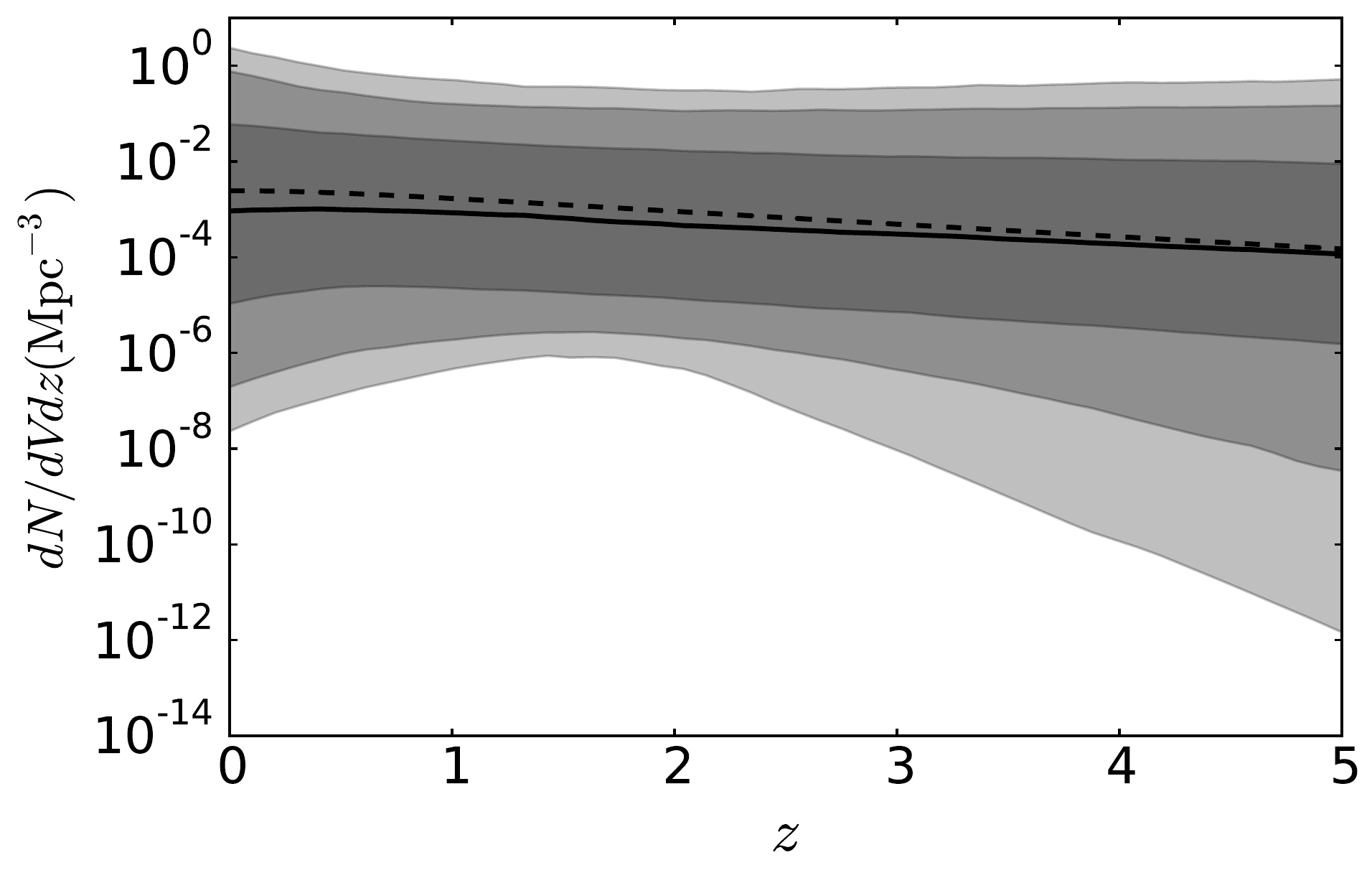}\\
\includegraphics[width=6.5cm]{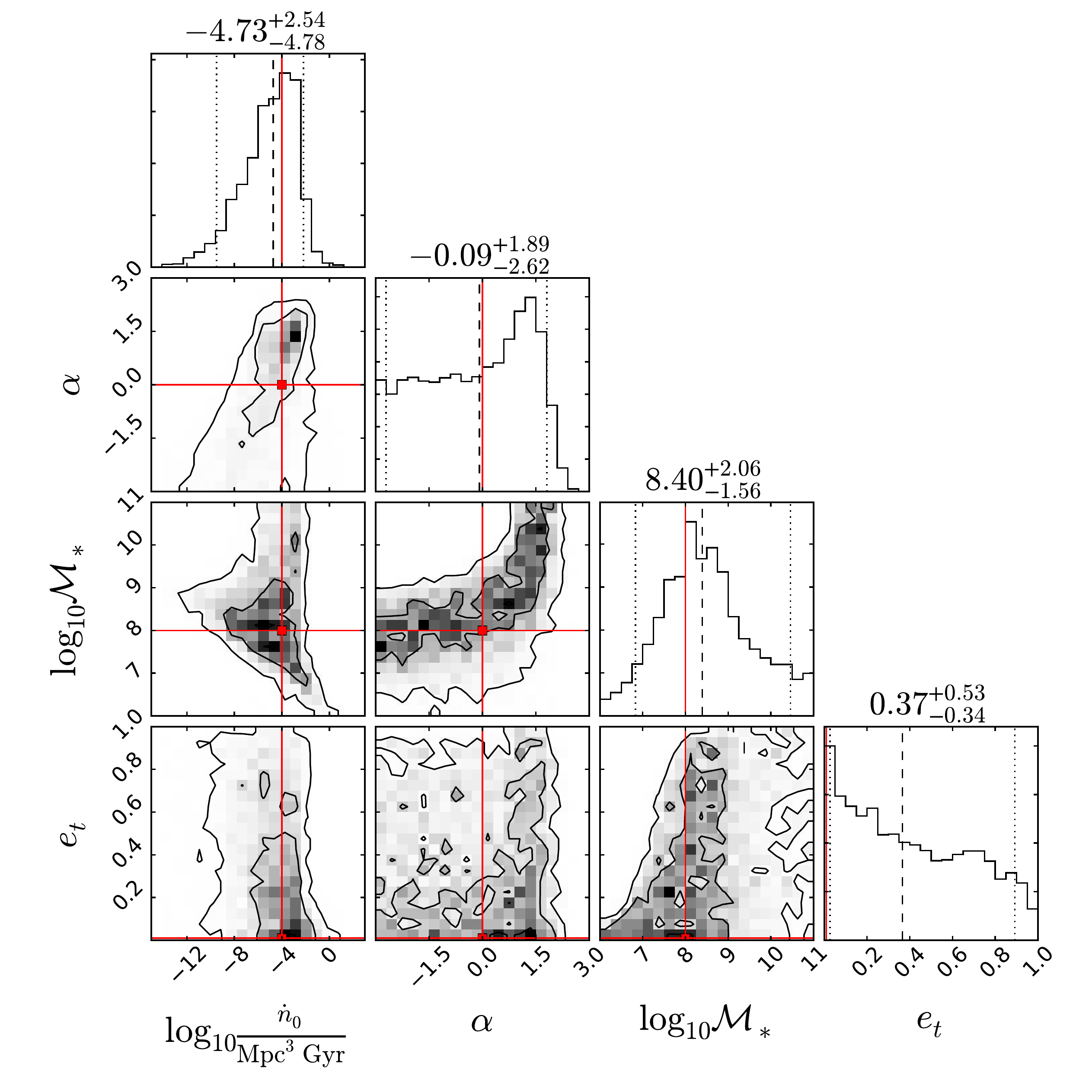}
\end{subfigure} \hspace{1cm}
\begin{subfigure}{0.425\textwidth}
\centering
$SKA20, \ e_t = 0.01$
\includegraphics[width=6.0cm]{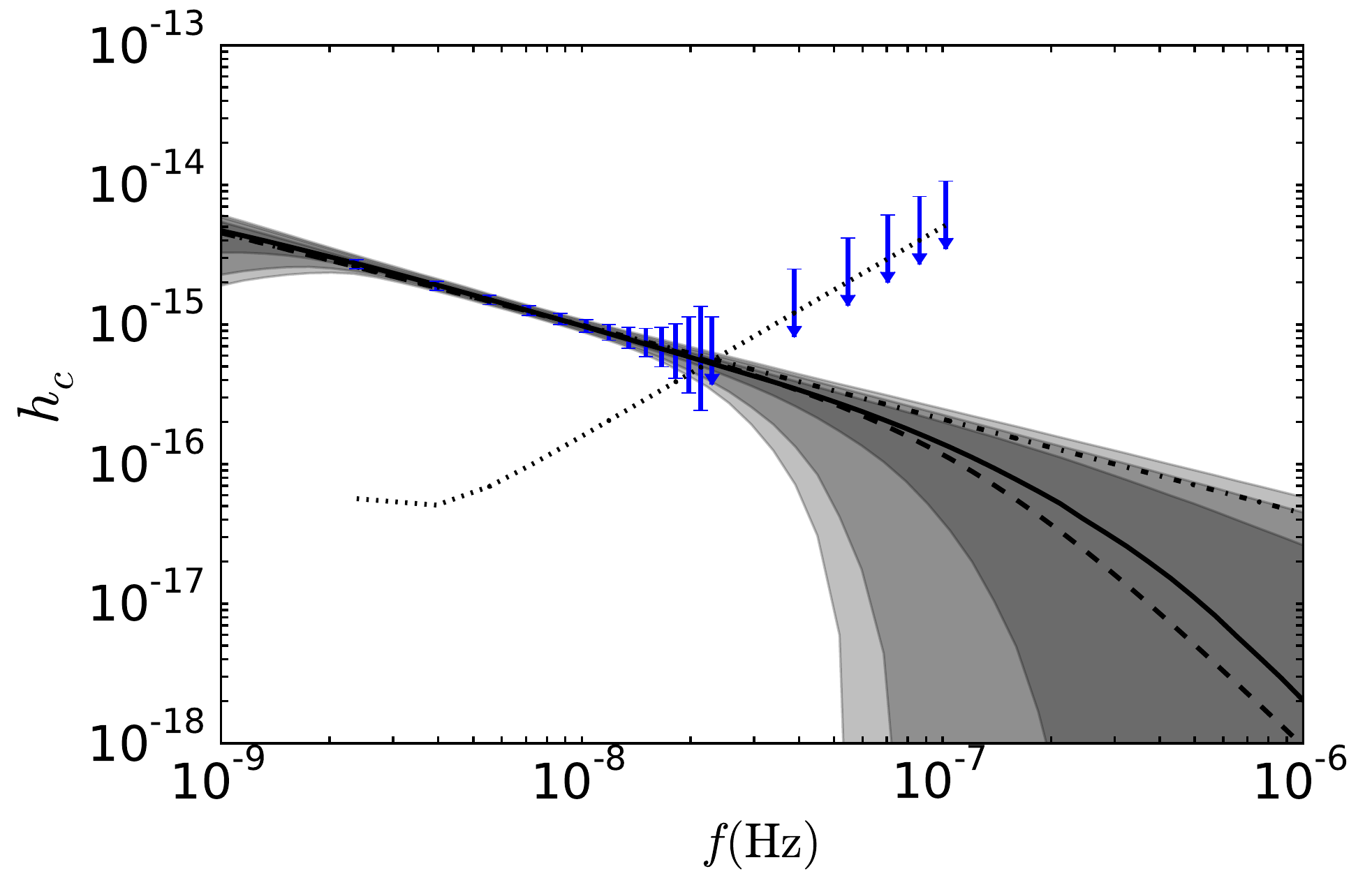}\\
\includegraphics[width=6.0cm]{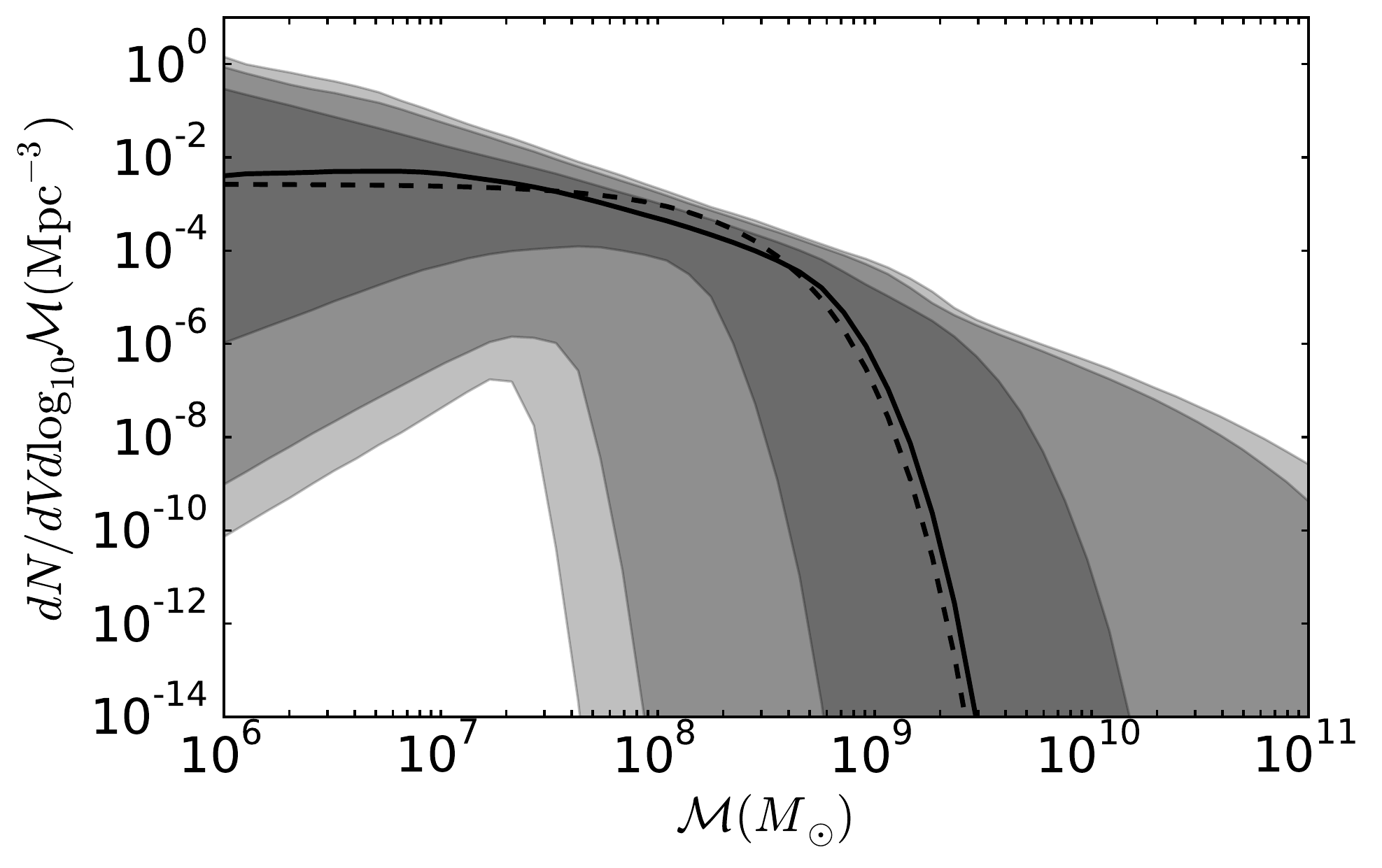}\\
\includegraphics[width=6.0cm]{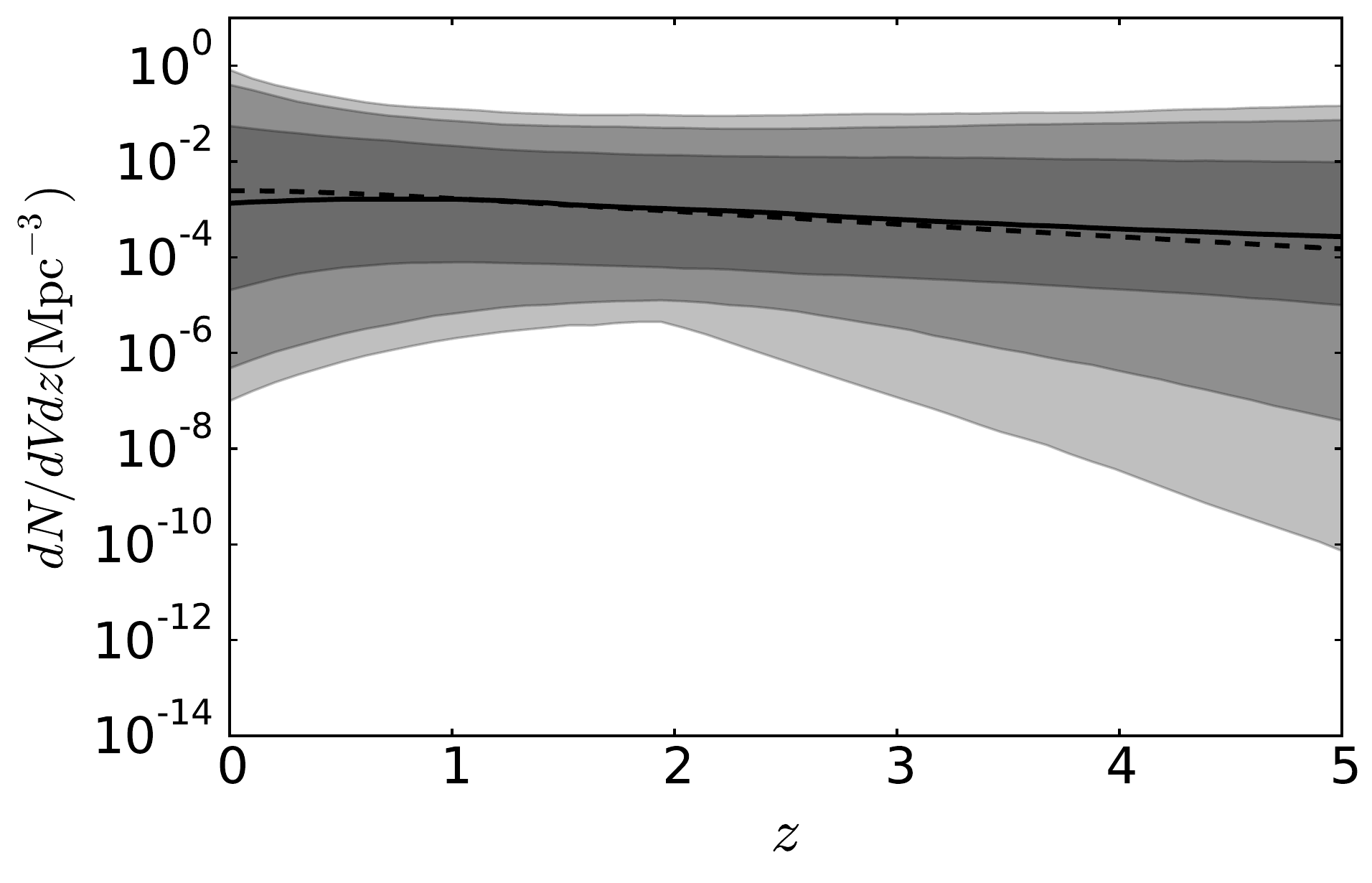}\\
\includegraphics[width=6.5cm]{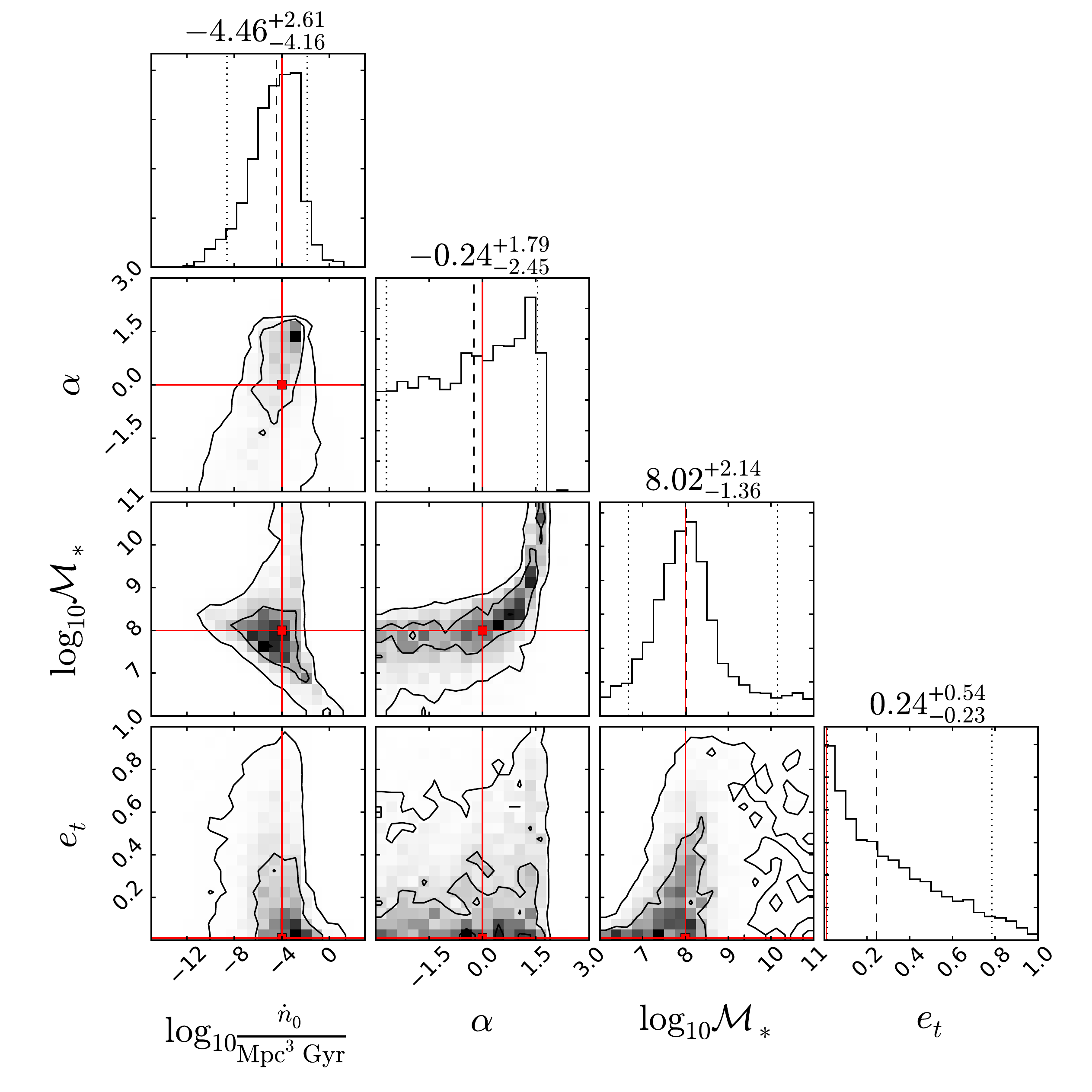}
\end{subfigure} \\
\caption{Implication of a PTA detection at a moderate (S/N$\approx$5, left column) and high (S/N$\approx$35, right column) significance, assuming a MBHB population with default mass function parameters and almost circular ($e_t=0.01$) eccentricity at decoupling. As in Fig. \ref{fig:upperlimit}, the posterior for the spectrum, mass and redshift functions (in descending order from the top) are shown as shaded areas, with the 68\%, 95\% and 99.7\% confidence regions indicated by progressively lighter shades of grey, and the solid black line marking the median of the posterior. In each of those panels, the dashed black line indicates the injected model. In the top panels the vertical blue bands indicate the 68\% confidence interval of the observed signal amplitude at each frequency bin, and the downward pointing arrows at higher frequency mark the 95\% upper limits. The dotted line is the nominal $1\sigma$ sensitivity of the considered PTA, as defined by equation \ref{eqhn}, where the contribution of $S_h$ to the noise has been omitted (see section \ref{sec:simobs} for details). The dot-dash black line shows the simulated spectrum assuming no drop in high mass sources. The lower triangular plots show the two-dimensional posteriors for each model parameter pairs, together with their marginalised distributions. The injected parameter values are marked by red solid lines and the black lines in each one dimensional distribution mark the median (dashed) and the central 90\% (dotted) of the posterior, along with the numerical values above each plot.}
\label{fig:detection_circ}
\end{figure*}
\begin{figure*}
\begin{subfigure}{0.425\textwidth}
\centering
$IPTA30, \ e_t = 0.9$
\includegraphics[width=6.0cm]{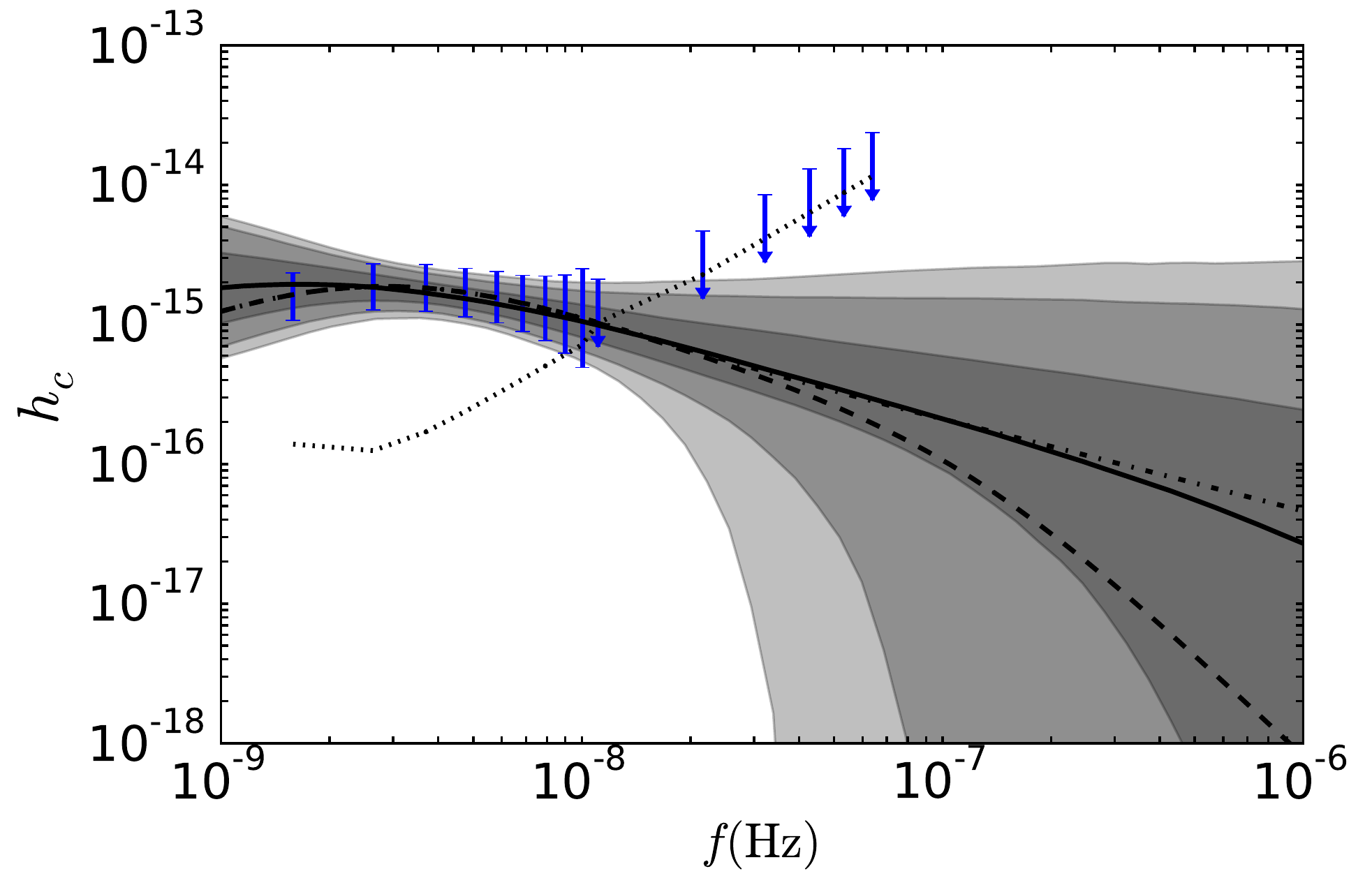}\\
\includegraphics[width=6.0cm]{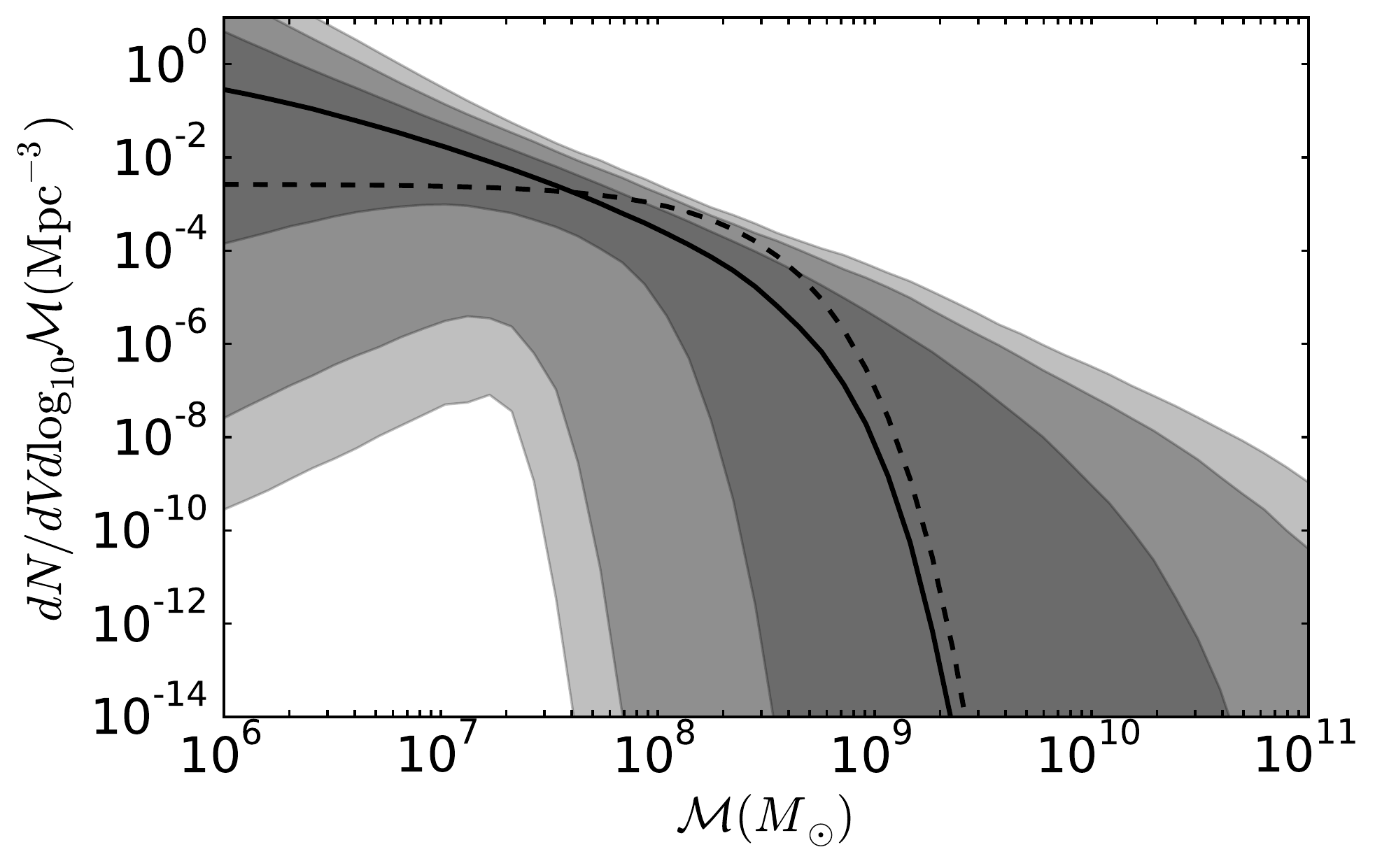}\\
\includegraphics[width=6.0cm]{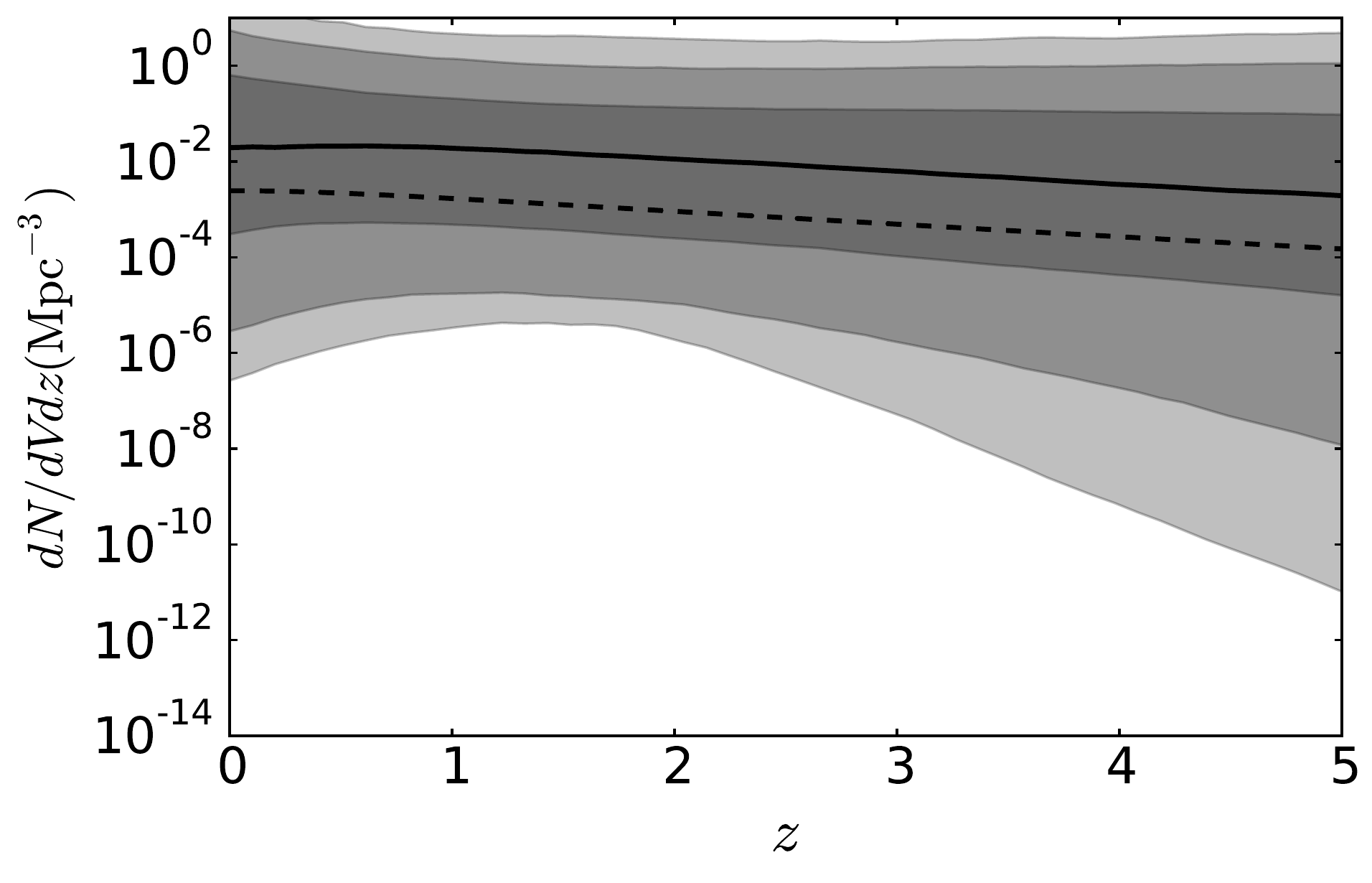}\\
\includegraphics[width=6.5cm]{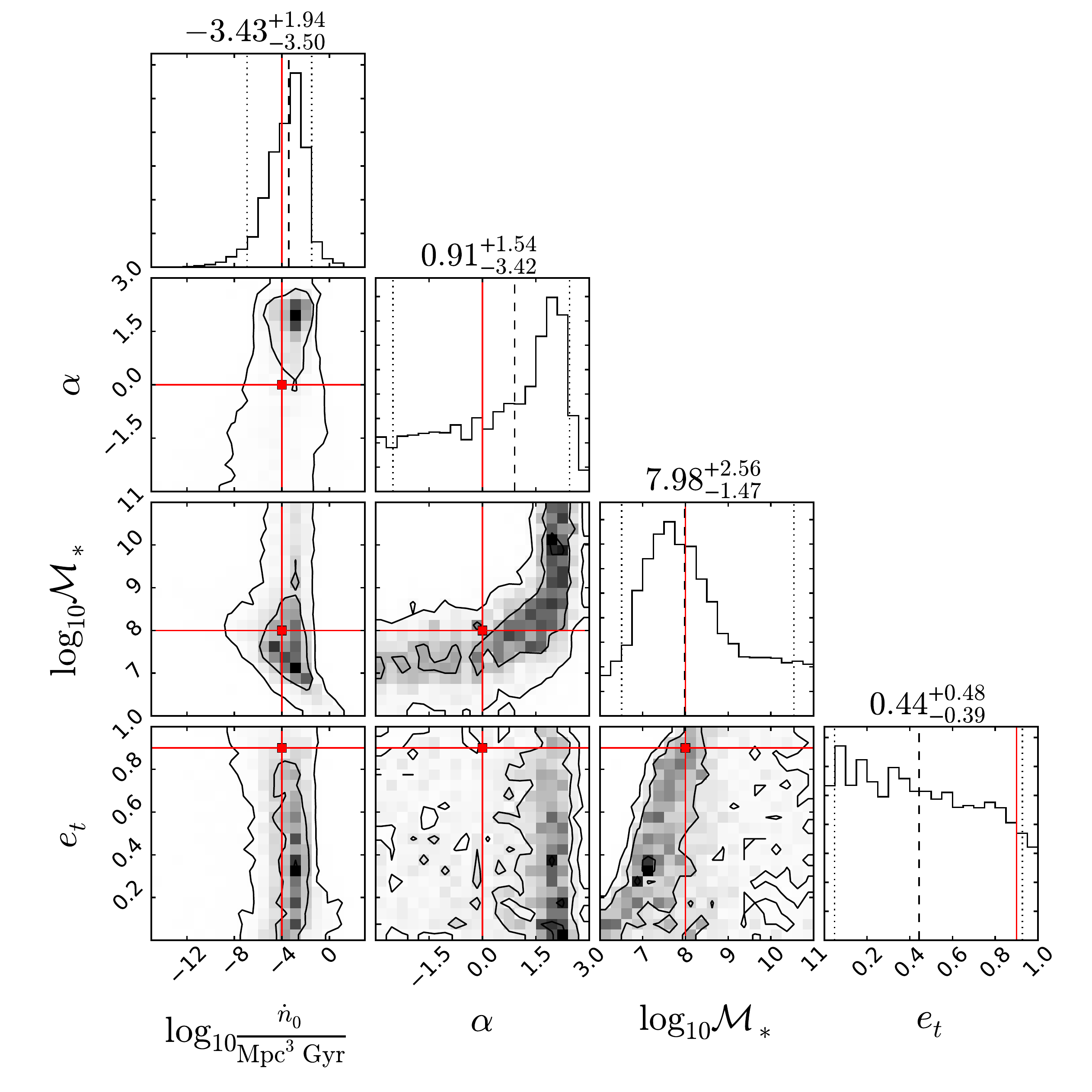}
\end{subfigure} \hspace{1cm}
\begin{subfigure}{0.425\textwidth}
\centering
$SKA20, \ e_t = 0.9$
\includegraphics[width=6.0cm]{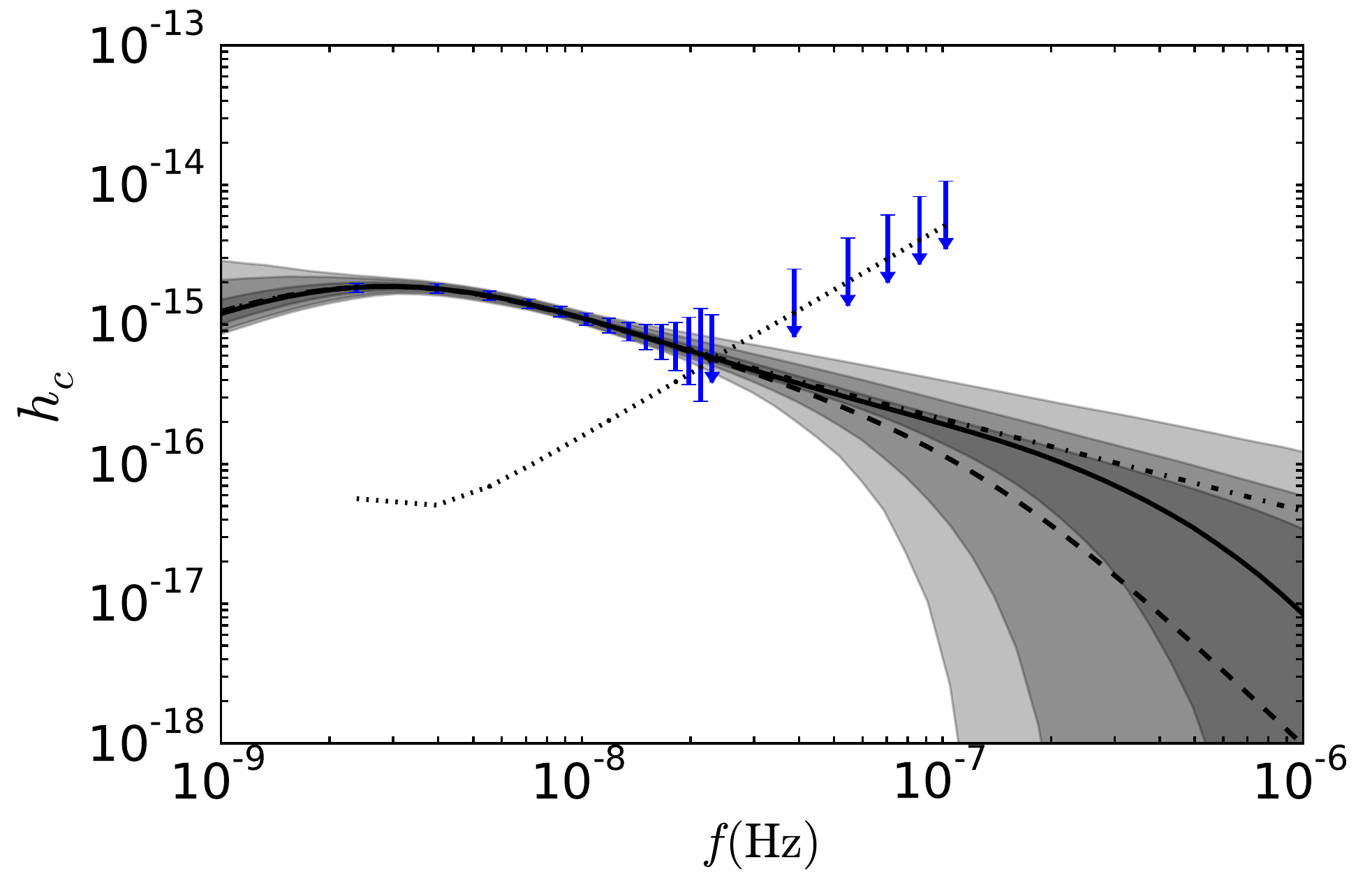}\\
\includegraphics[width=6.0cm]{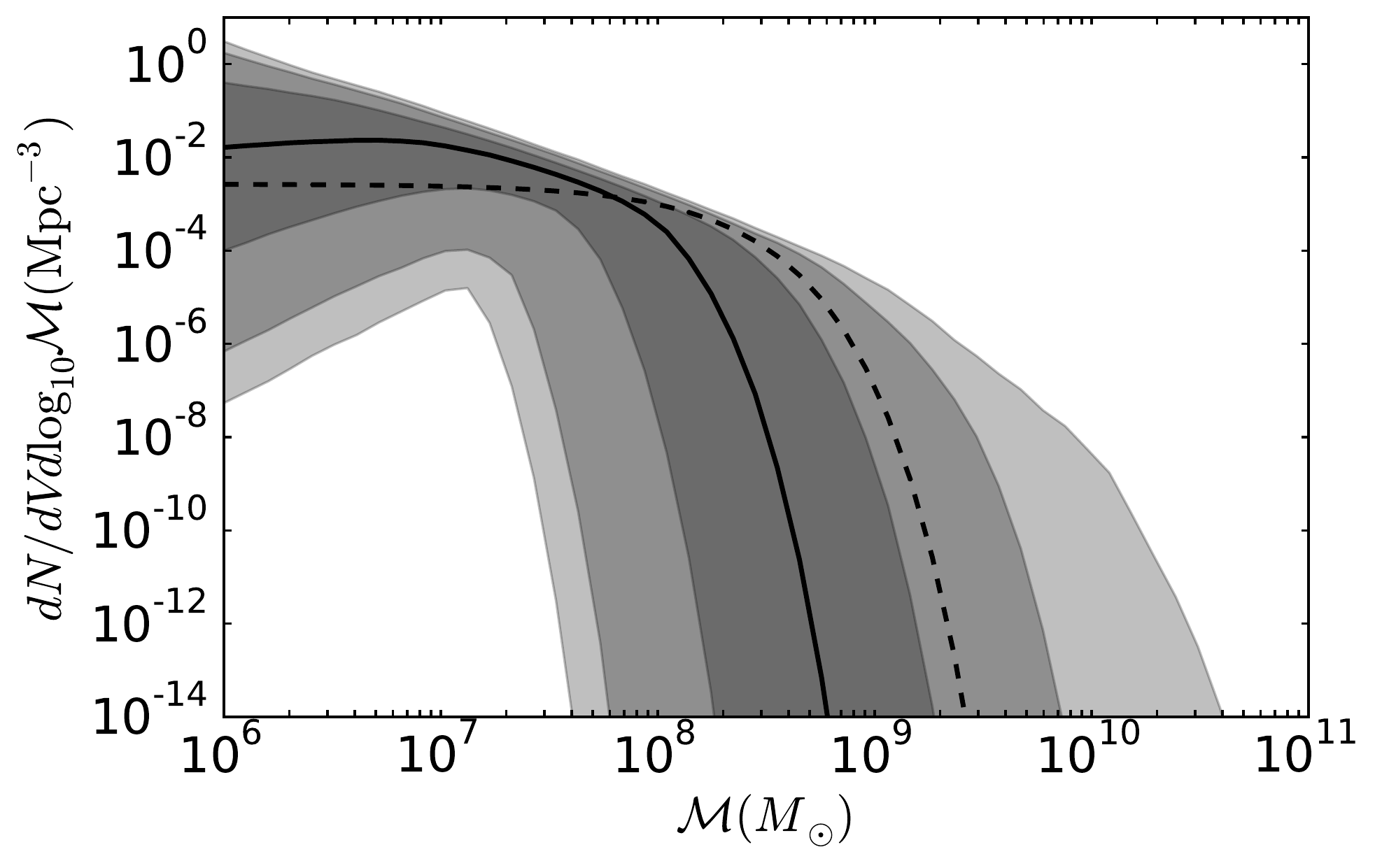}\\
\includegraphics[width=6.0cm]{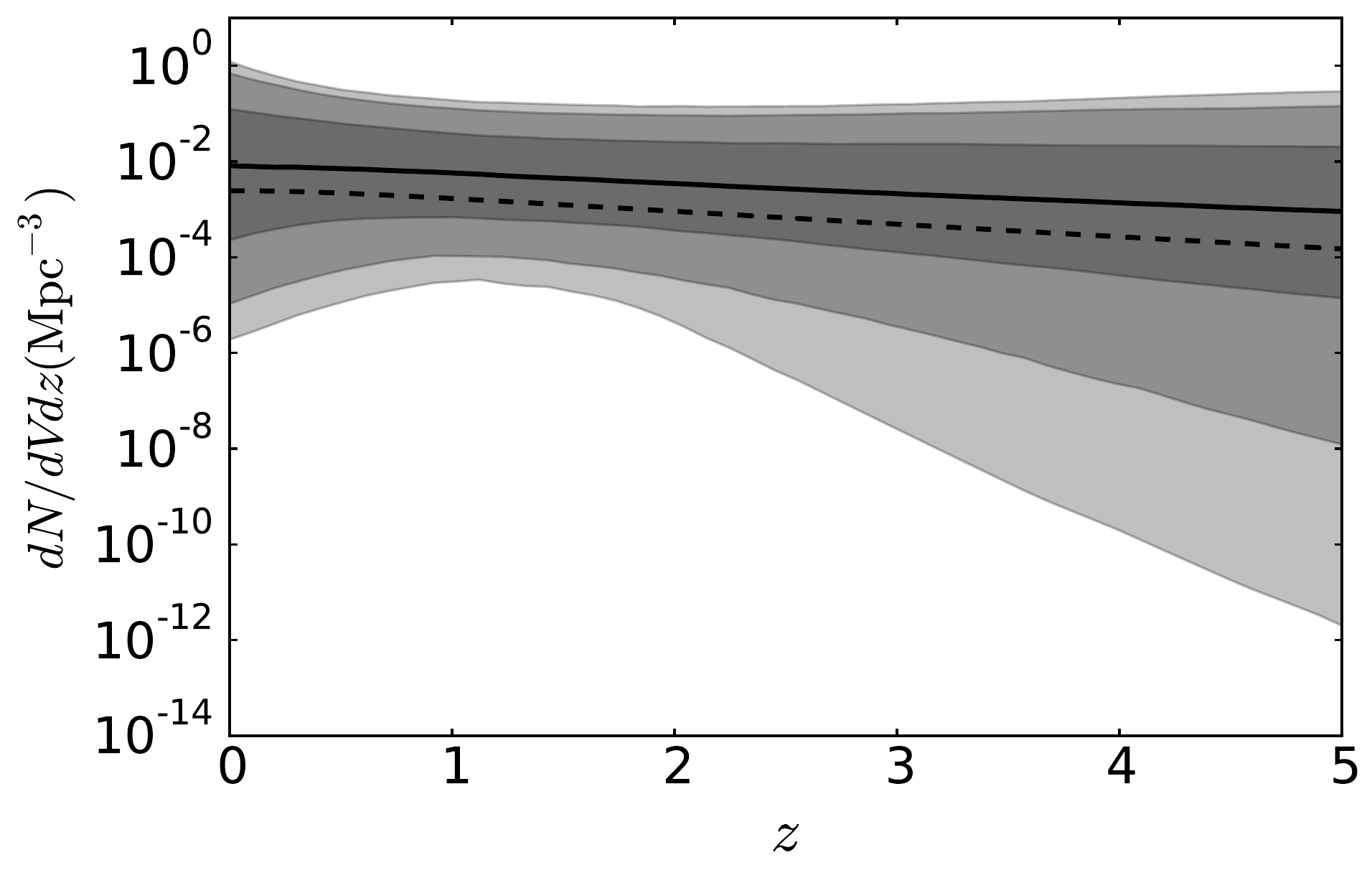}\\
\includegraphics[width=6.5cm]{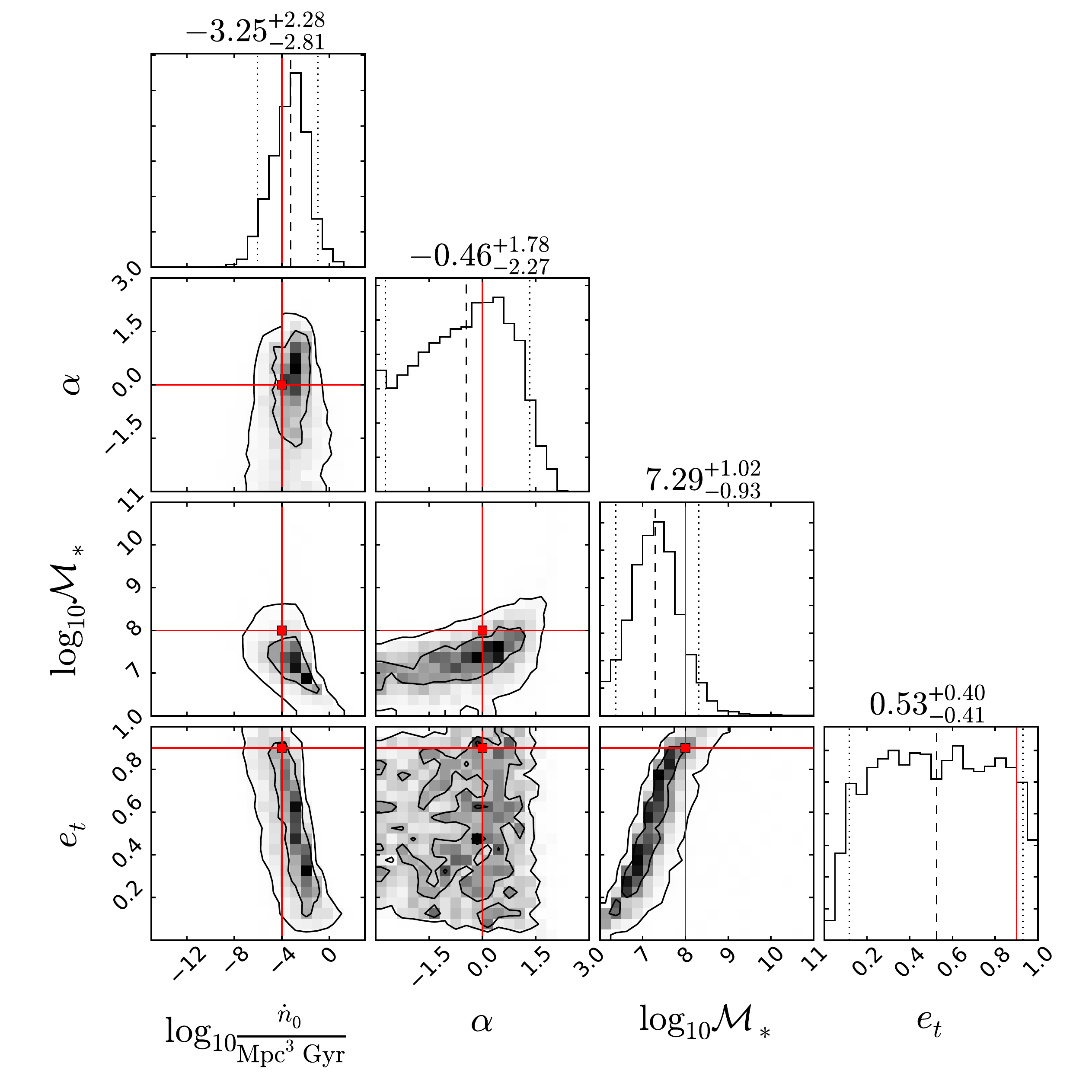}
\end{subfigure} \\
\caption{Same as figure \ref{fig:detection_circ} but assuming decoupling eccentricity of $e_t=0.9$.}
\label{fig:detection_ecc}
\end{figure*}

We turn now to the implication of a future PTA detection. We discuss two distinct MBHB populations corresponding to our default mass function model (with parameters given in section \ref{sec:simsetup}) and defined by decoupling eccentricity $e_t=0.01$ (circular case) and $e_t=0.9$ (eccentric case).

\subsubsection{Circular case}
Results for the circular case are shown in Fig. \ref{fig:detection_circ} to which we refer in the following discussion. In the {\it IPTA30} scenario (left column), the signal is detected in the lowest eight frequency bins, with total S/N$\approx6$. At $f<10$ nHz the spectrum is well constrained (upper panel), and the reconstructed MBHB mass function and redshift distribution (central panels) are consistent with the injected values. Note, however, that astrophysical constraints are quite poor; even around $\Mc=3\times 10^8\Msol$, where the mass function is best constrained, the 68\% confidence interval spans about two order of magnitude, and so does the high mass cut-off. The triangle plot in the lower panel provides more insight into the reconstruction of the model parameters. In general, the posteriors of all the parameters are consistent with the injected values, however the distributions are fairly broad and the contour plots unveil several correlations among model parameters, the most important of which will be investigated later on.

The situation quantitatively improves, but is qualitatively unaltered, in the {\it SKA20} scenario, shown in the right column. Here the signal is detected in 13 frequency bins, with a total S/N$\approx 35$. The $h_c$ spectrum is extremely well reconstructed up to $20$nHz and the median of the recovered mass and redshift functions match the injected ones almost exactly (central panel); uncertainties are still large though, and the posterior distributions of the model parameters improve only marginally. The characteristic mass scale ${\cal M}_*$ is slightly better constrained and, compared to the {\it IPTA30} case, there is a stronger preference for circular binaries, although higher eccentricity cannot be ruled out.

\subsubsection{Eccentric case: parameter degeneracies}

The eccentric case is shown in Fig. \ref{fig:detection_ecc}. Again, in the {\it IPTA30} scenario (left column panels) the signal is detected in the nine lowest frequency bins, with total S/N$\approx5$. The recovered GW spectrum is consistent with the injected one, but errors are large and the shape can be hardly determined. The triangle plot in the lower-left panel shows that it is difficult to recover model parameters. Posteriors are consistent with injected values, but the distributions are hardly informative.

  Moving to the {\it SKA20} case (right column panels), we see a clear improvement on the reconstruction of the spectrum (upper panel), but the preferred mass function appears quite offset with respect to the original injection (second panel from the top). Posterior distributions in the triangle plot (lower panel) are now more informative and reveal more defined degeneracies. Particularly interesting is the $\int$-shaped posterior in the $e_t-\Mstar$ panel (already visible in the {\it IPTA30} case). The degeneracy stems from the mass dependence of the decoupling frequency in equation (\ref{eq:fdec}), i.e. from the fact that more massive MBHBs decouple at lower frequencies than lighter ones. In fact, for a given eccentric MBHB, the peak of the GW spectrum occurs at a frequency $f_p={\cal F}(e_t)f_d$ (see equation 13 in PaperI), where ${\cal F}(e_t)$ is a monotonically increasing function of $e_t$. This means that, if we observe a turnover in the GWB at a given $\bar{f}$, there is an ambiguity in the determination of the decoupling eccentricity of the MBHB population. The signal can be dominated by lighter MBHB decoupling at higher $f_d$ with lower $e_t$, or by heavier MBHB decoupling at lower $f_d$ with higher $e_t$, giving rise to the $\int$-shaped contour in the $e_t-\log_{10}\Mstar$ plane. Lighter black holes require a higher $\ndot$ to produce the observed signal level, however this is still well within the assumed prior. In practice, the detection of a turnover in the GWB, guarantees that MBHBs have some eccentricity at decoupling (which in our models always occur below the observable PTA frequency window), however cannot inform us on the value of their eccentricity, unless independent information on the MBHB mass function becomes available. This causes the peculiar shape of the $e_t$ posterior seen in the lower-right panel of Fig. \ref{fig:detection_ecc}, in which the posterior is quite flat down to $e_t\approx 0.1$ and has a sharp decline disfavouring circular binaries.

\subsection{Breaking degeneracies: the importance of detection at high frequencies}

\begin{figure*}
\begin{subfigure}{0.425\textwidth}
\centering
$Ideal, \ e_t = 0.01$
\includegraphics[width=6.0cm]{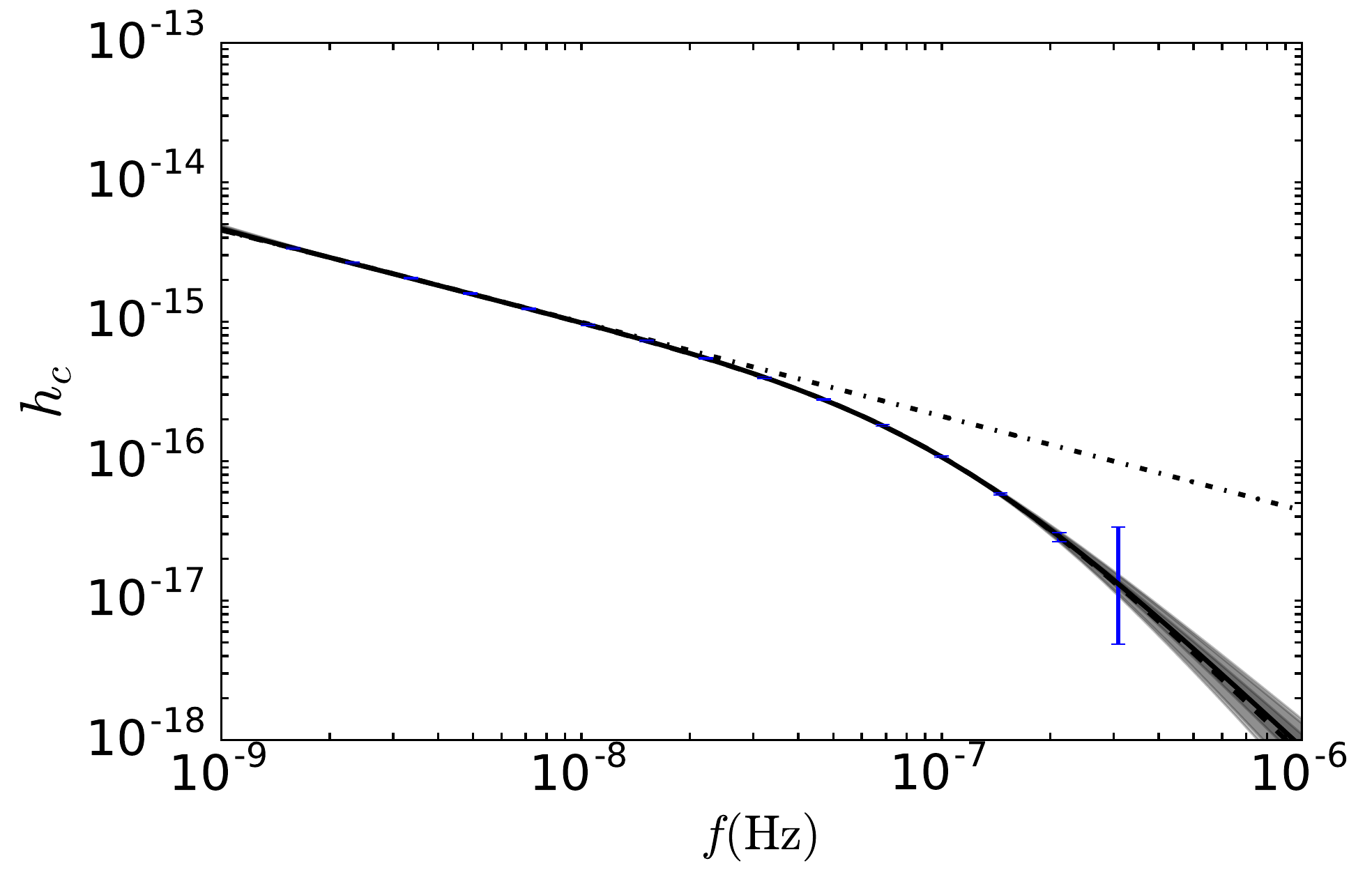}\\
\includegraphics[width=6.0cm]{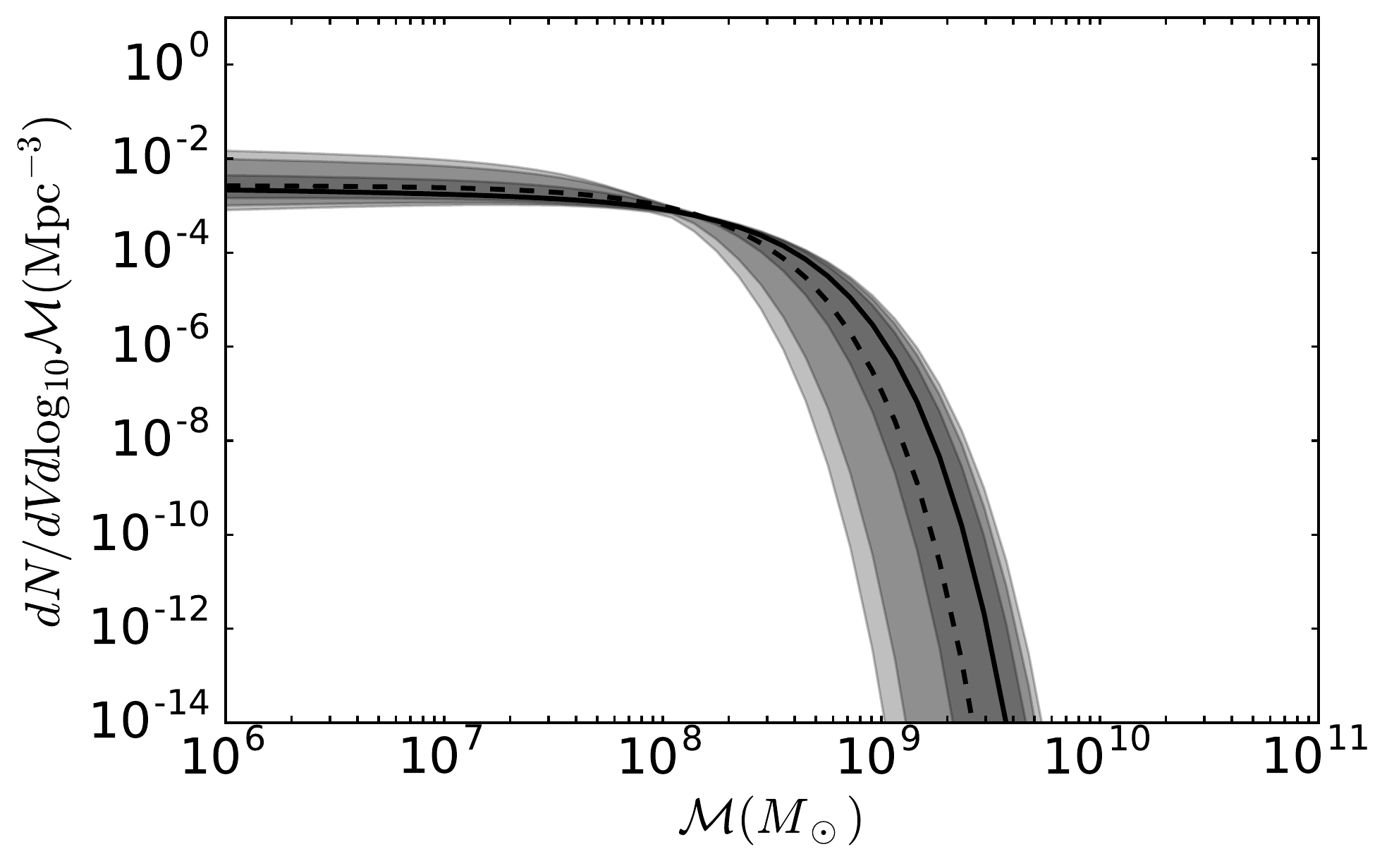}\\
\includegraphics[width=6.0cm]{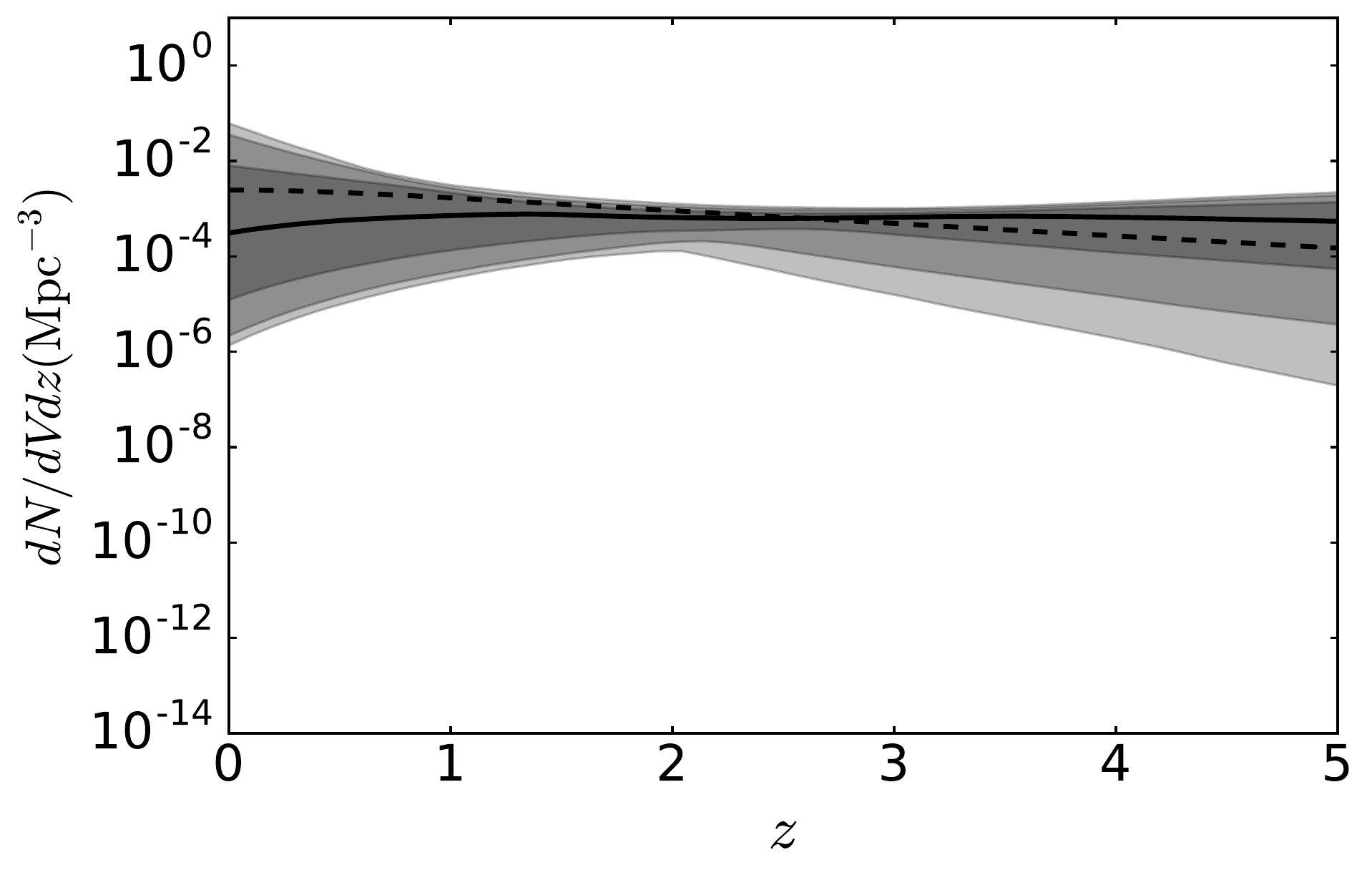}\\
\includegraphics[width=6.5cm]{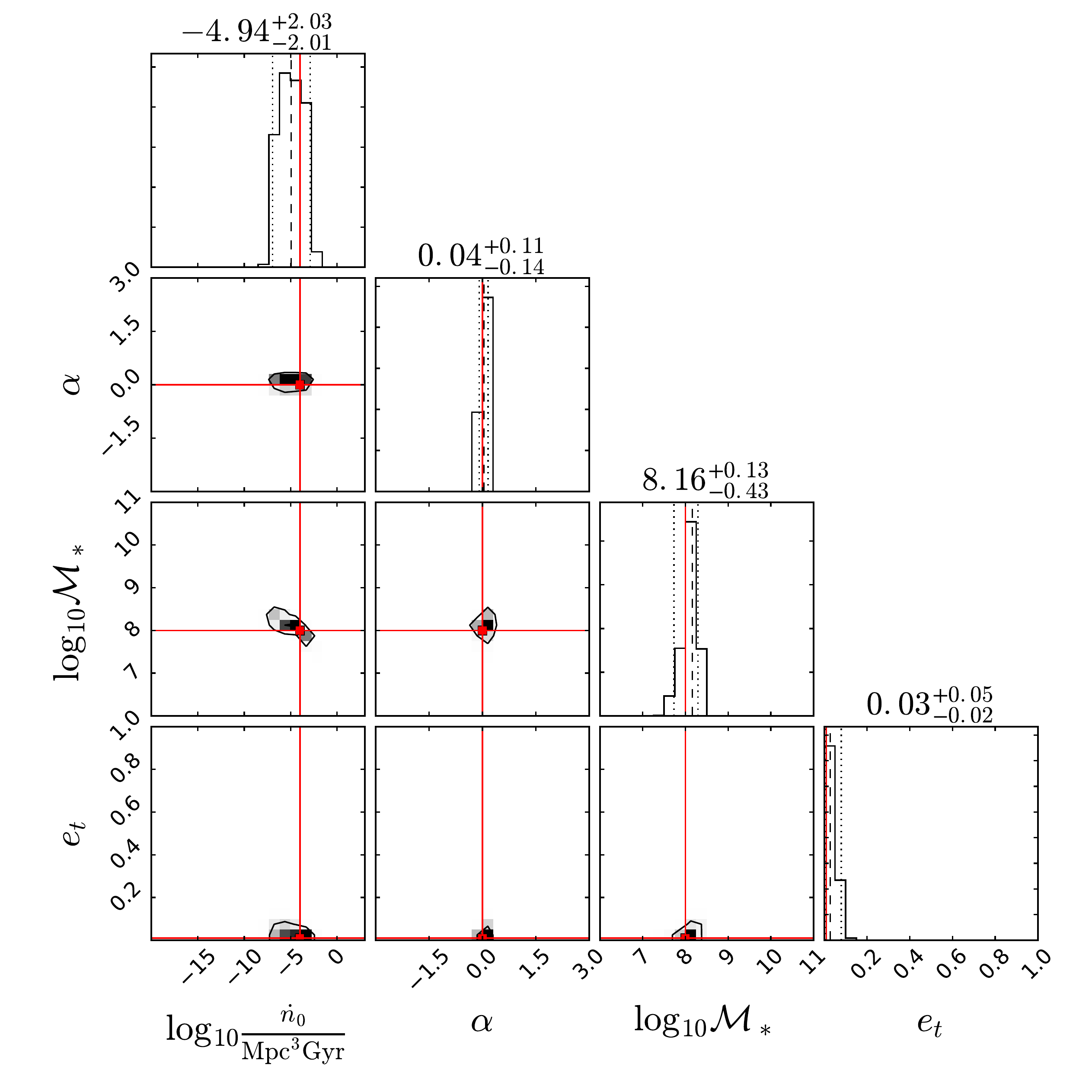}
\end{subfigure} \hspace{1cm}
\begin{subfigure}{0.425\textwidth}
\centering
$Ideal, \ e_t = 0.9$
\includegraphics[width=6.0cm]{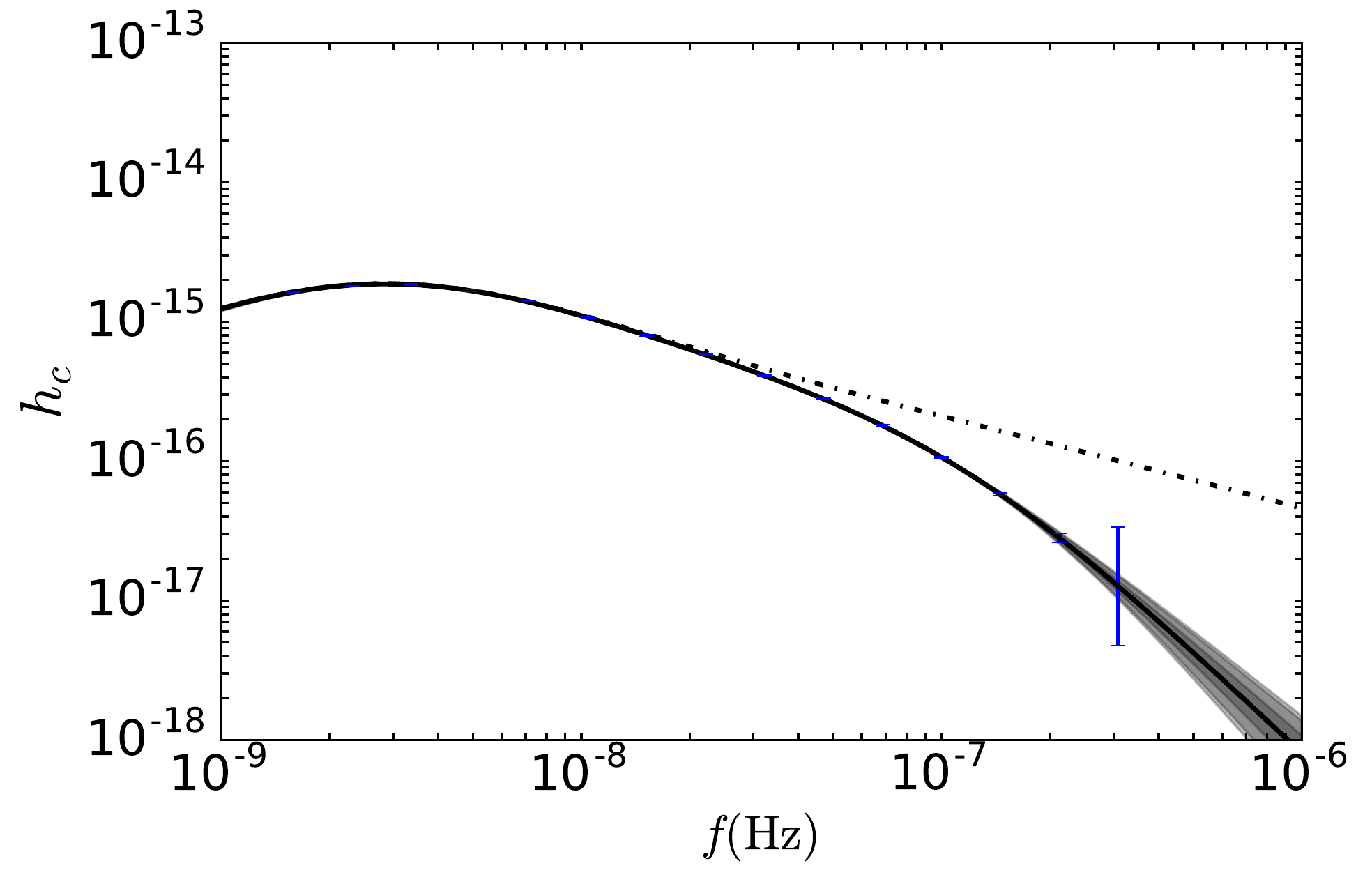}\\
\includegraphics[width=6.0cm]{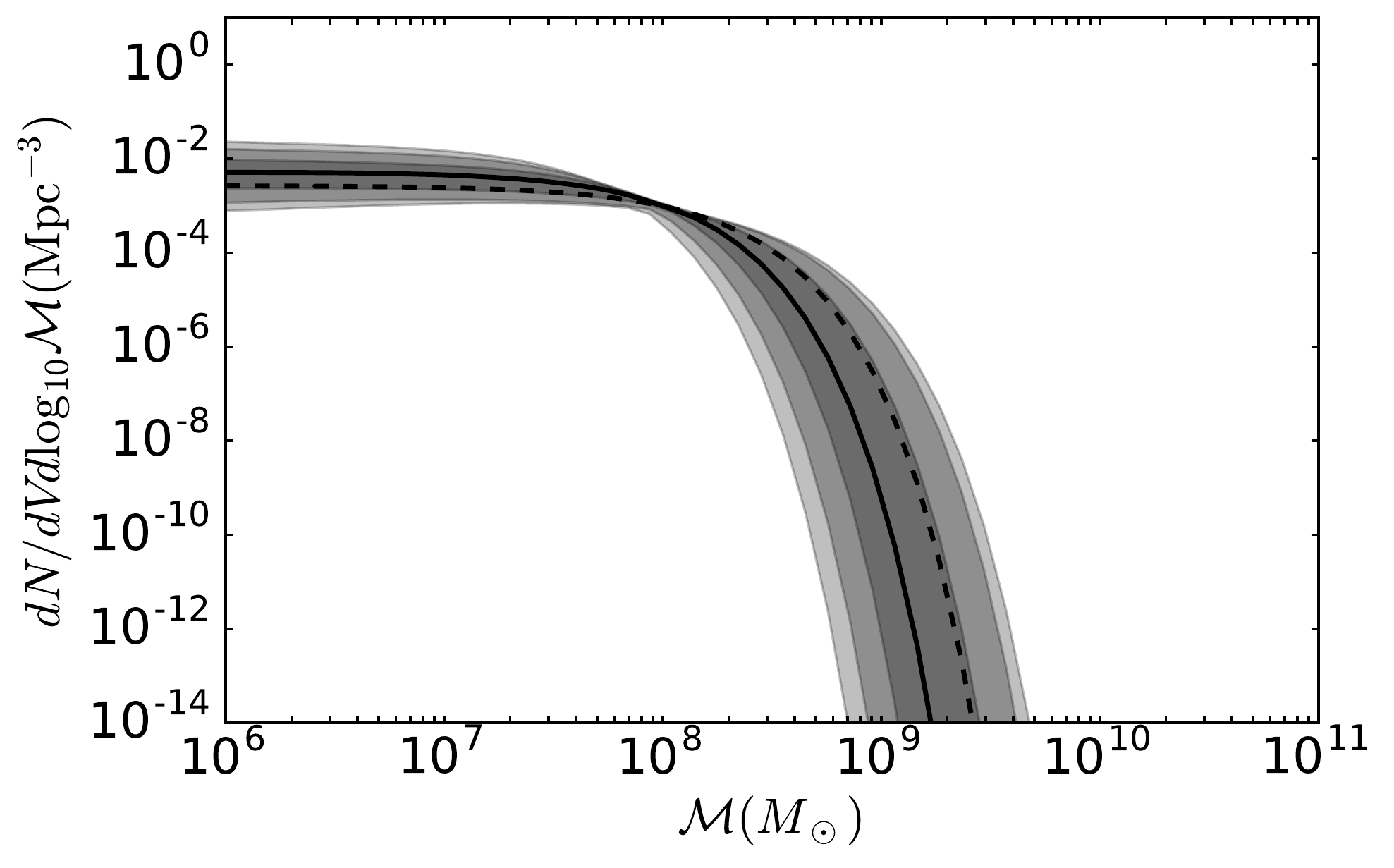}\\
\includegraphics[width=6.0cm]{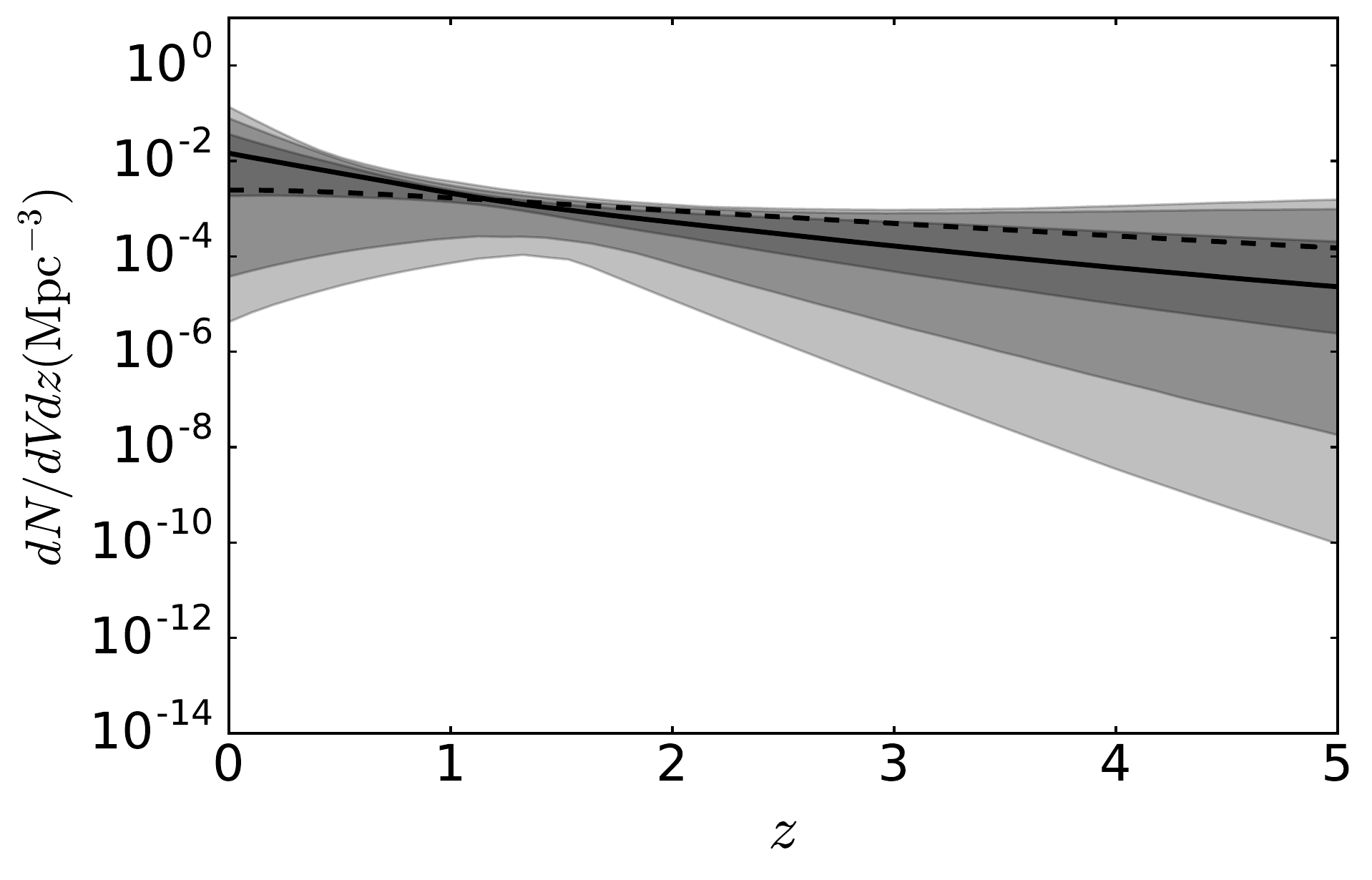}\\
\includegraphics[width=6.5cm]{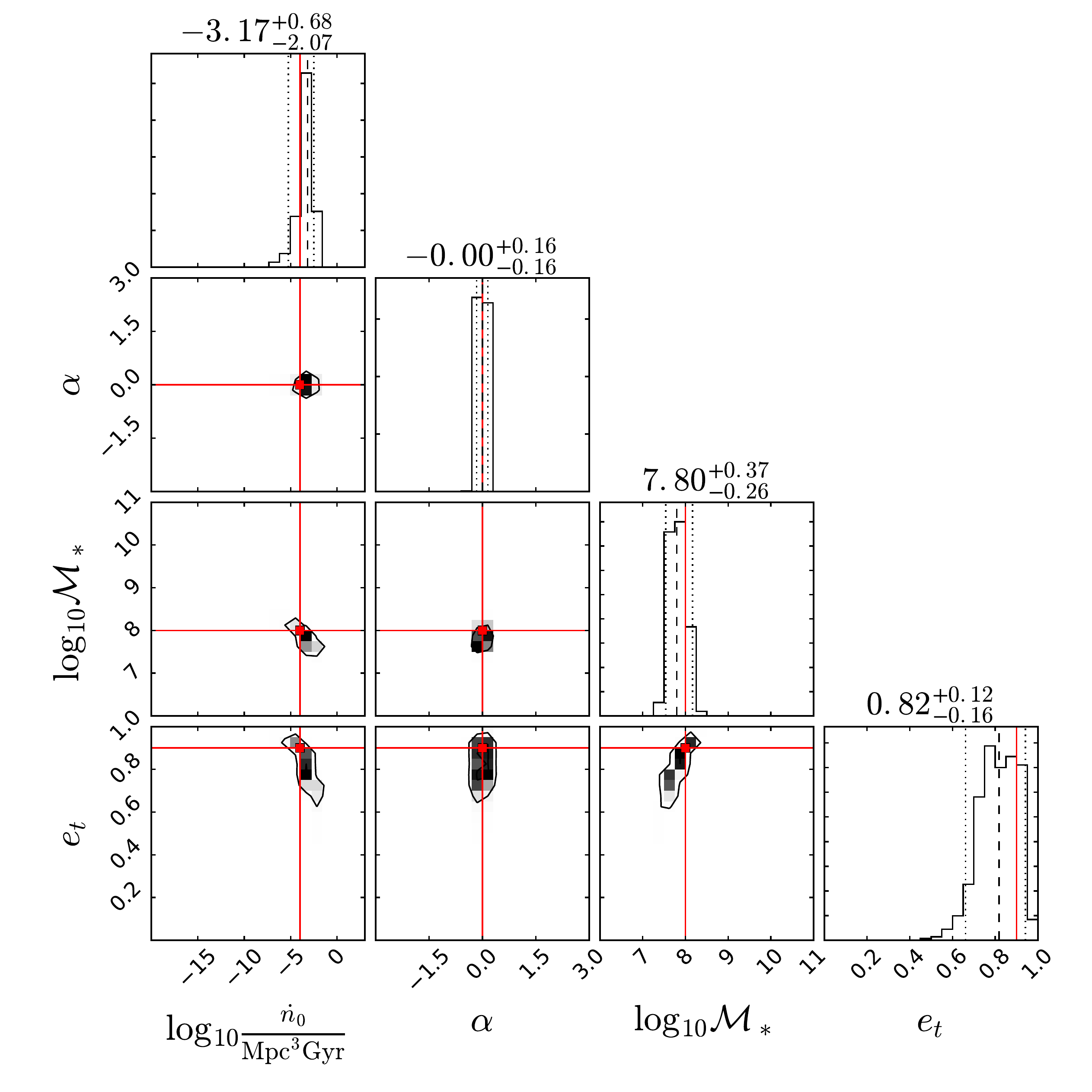}
\end{subfigure} \\
\caption{Implication of an ideal detection with 500 MSPs timed at sub-ns precision for 30 years. The injected model has default parameters with $e_t=0.01$ (left column) and $e_t=0.9$ (right column). Panel sequence and style as in Fig. \ref{fig:detection_circ}.}
\label{fig:cornerplot_ideal}
\end{figure*}
\begin{figure*}
\begin{subfigure}{0.425\textwidth}
\centering
$IPTA, \ e_t = 0.90$
\includegraphics[width=6.5cm]{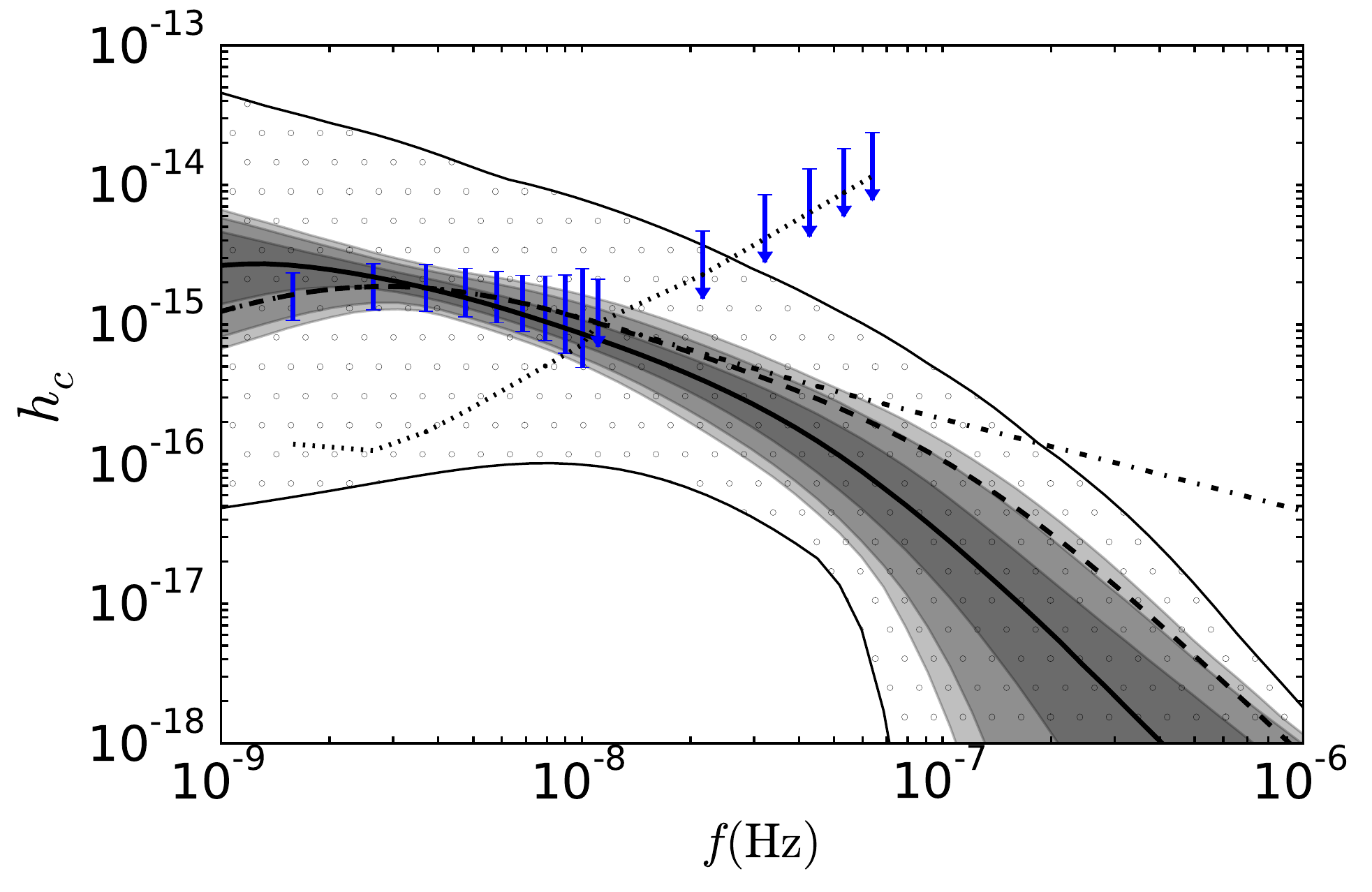}\\
\includegraphics[width=6.5cm]{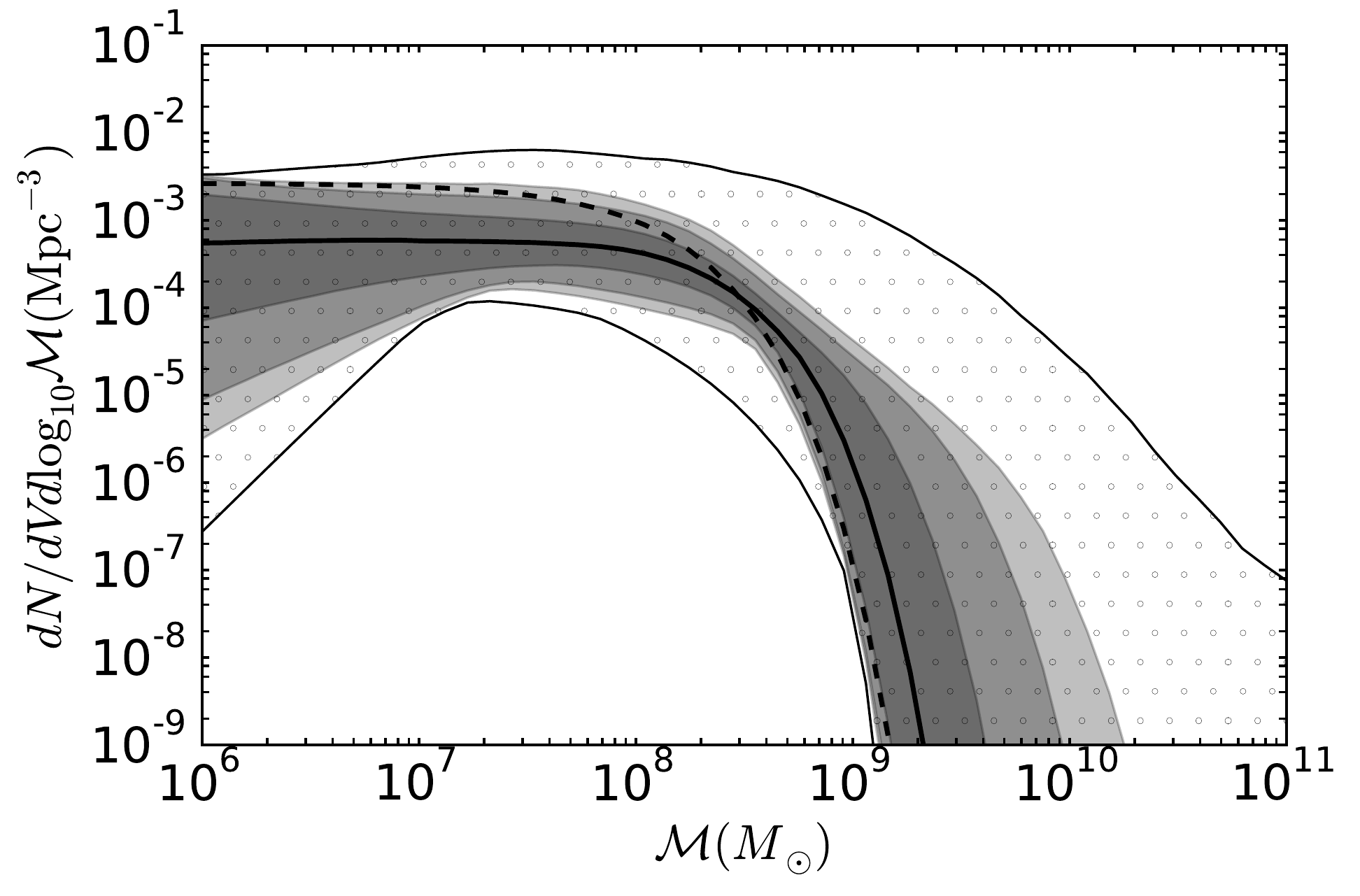}\\
\includegraphics[width=6.5cm]{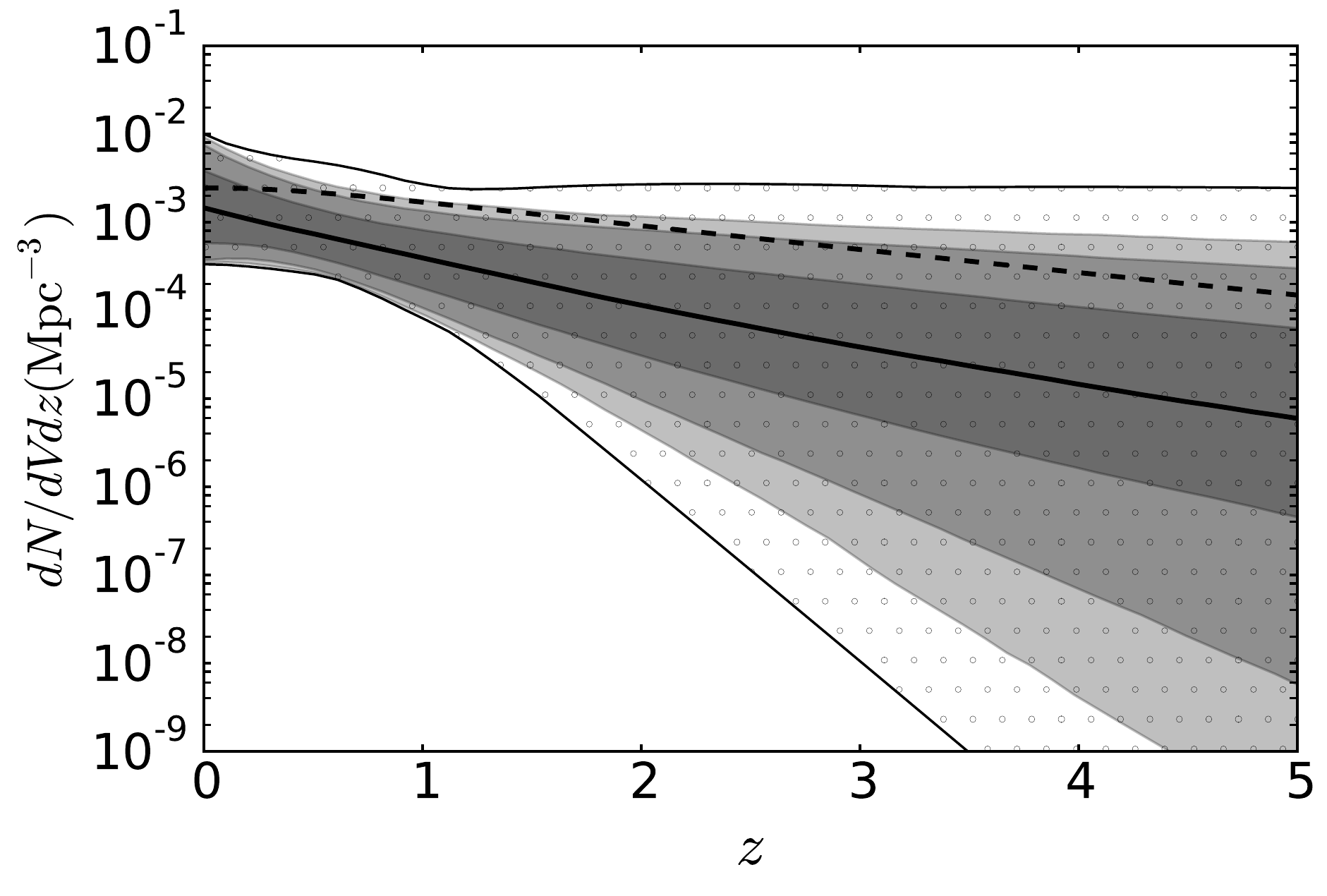}\\
\includegraphics[width=8.0cm]{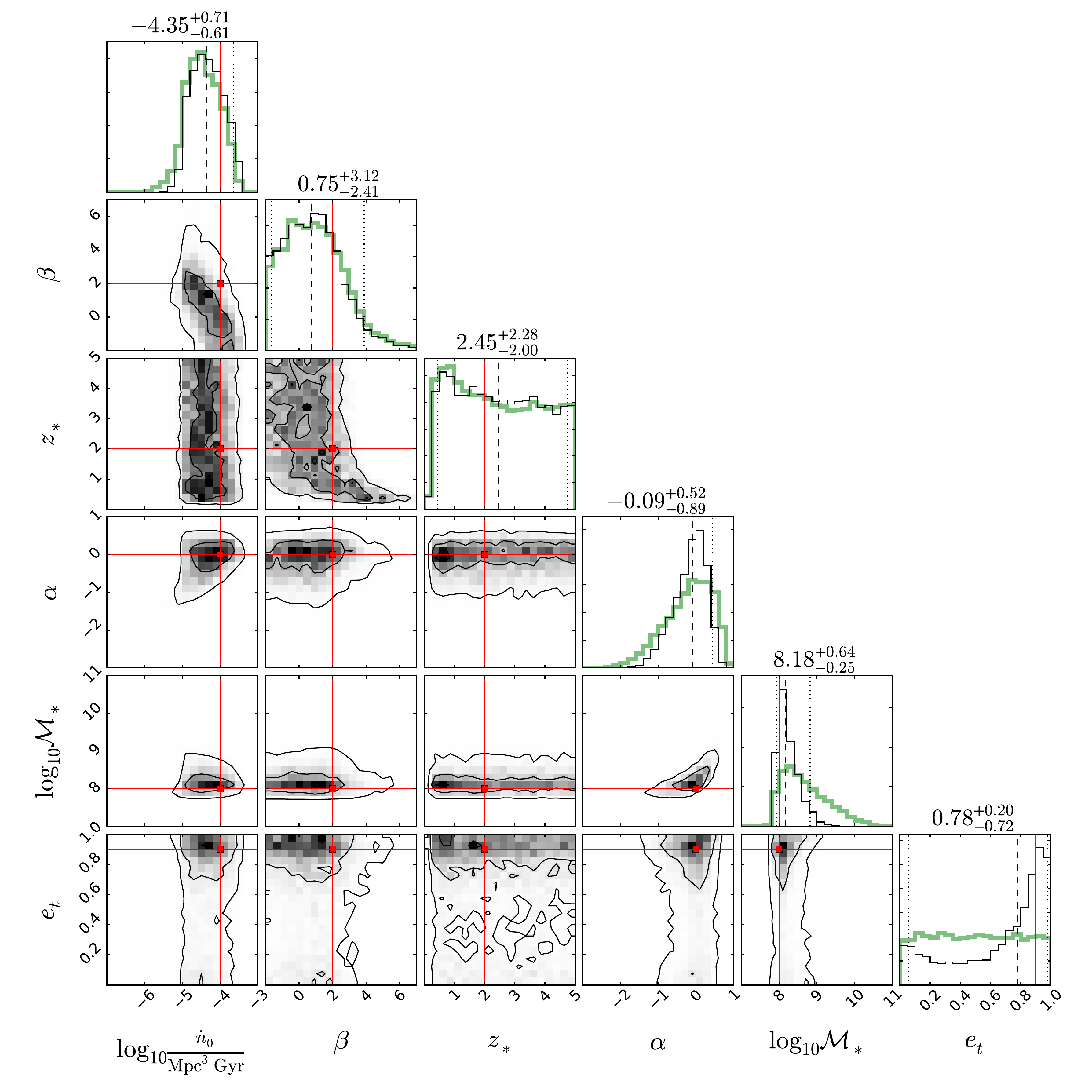}
\end{subfigure} \hspace{.5cm}
\begin{subfigure}{0.425\textwidth}
\centering
$Ideal, \ e_t = 0.90$
\includegraphics[width=6.5cm]{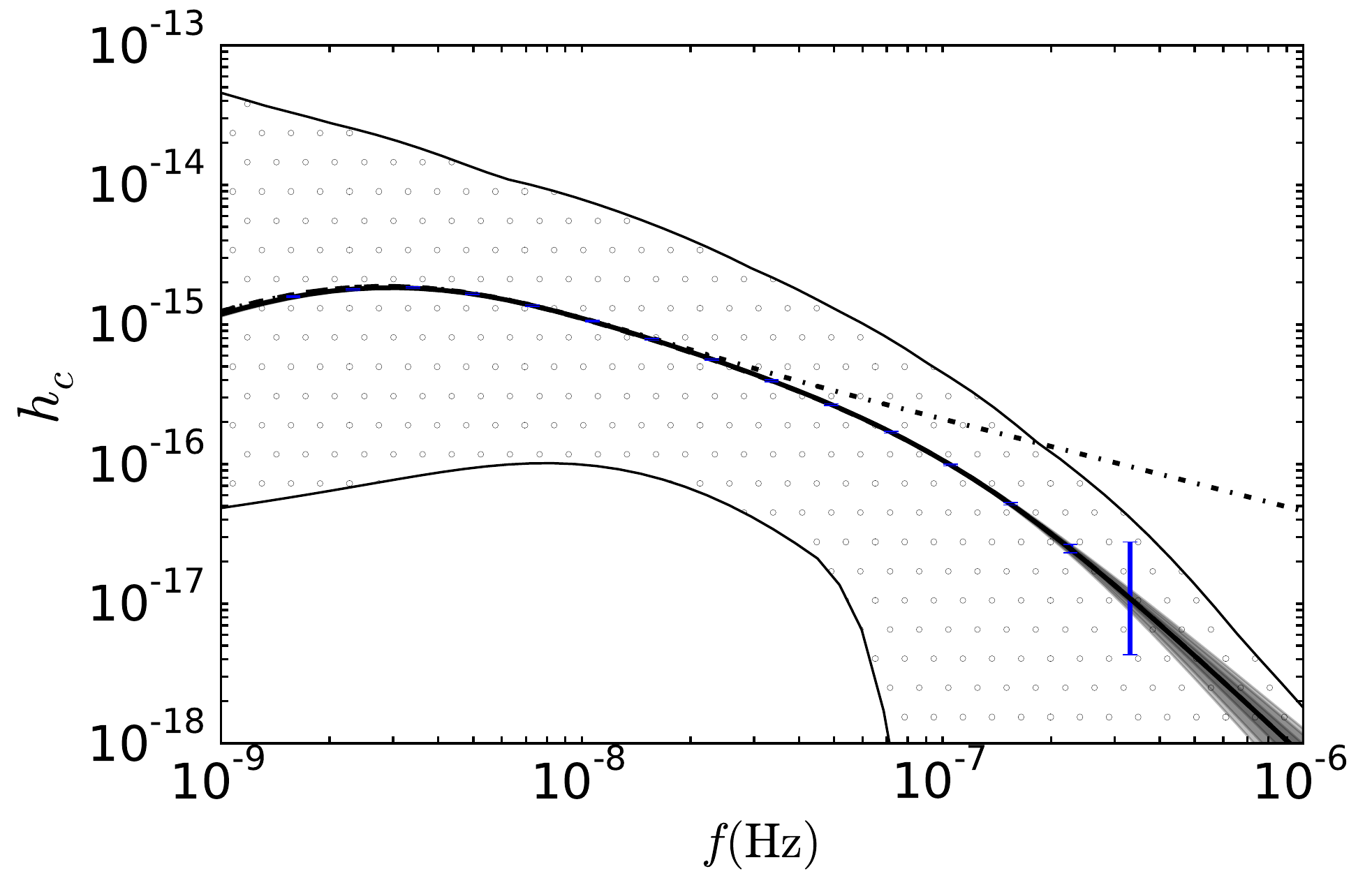}\\
\includegraphics[width=6.5cm]{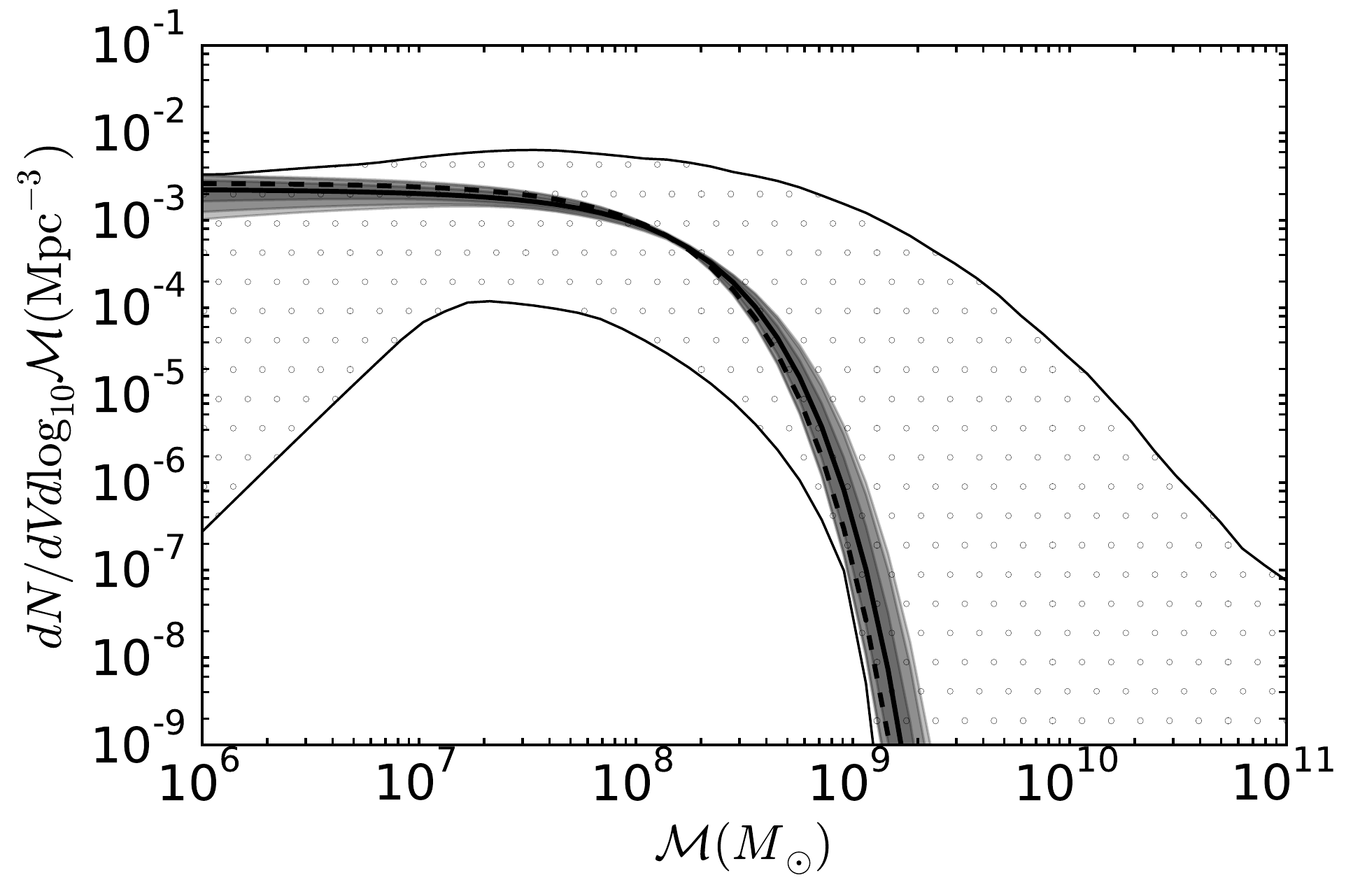}\\
\includegraphics[width=6.5cm]{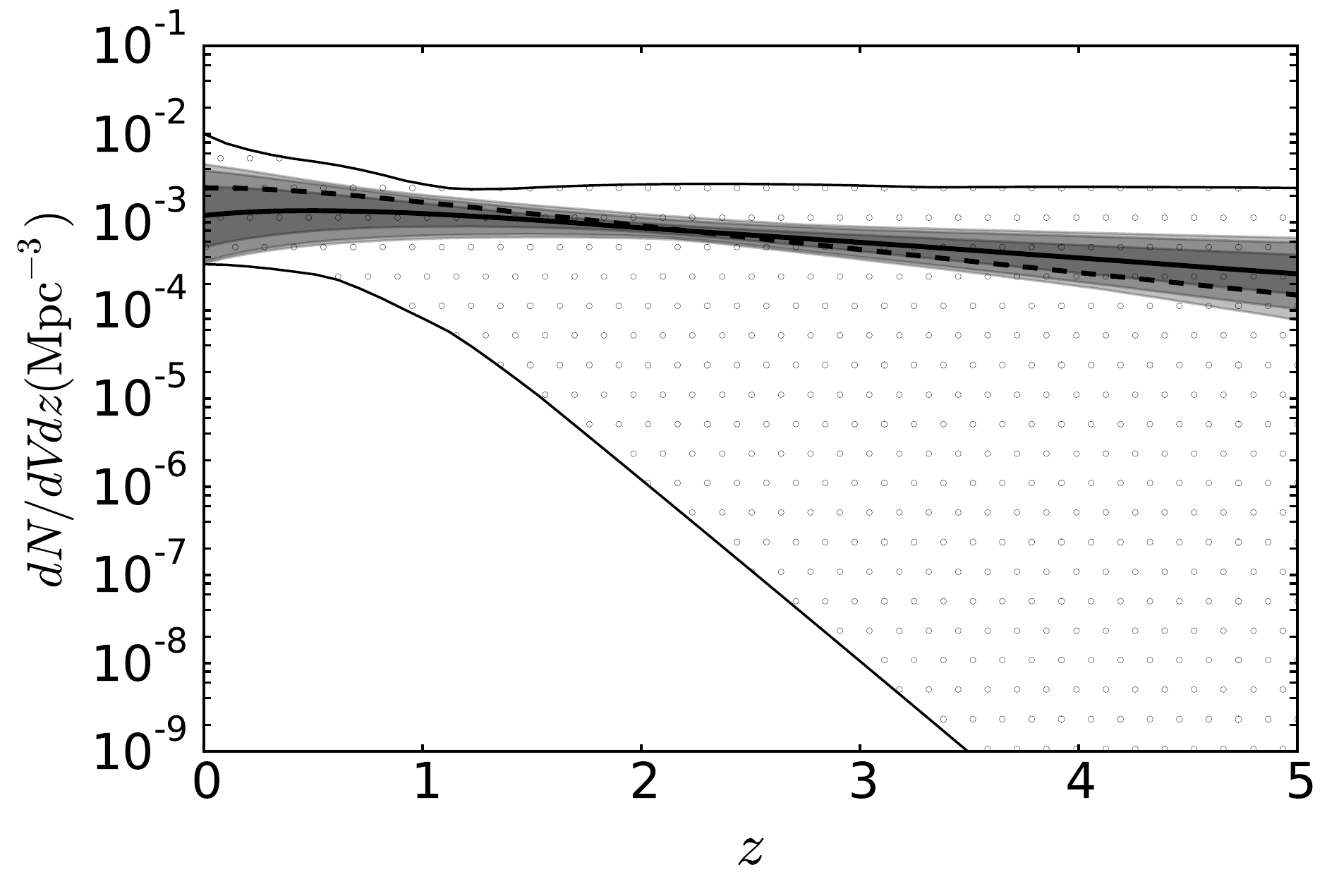}\\
\includegraphics[width=8.0cm]{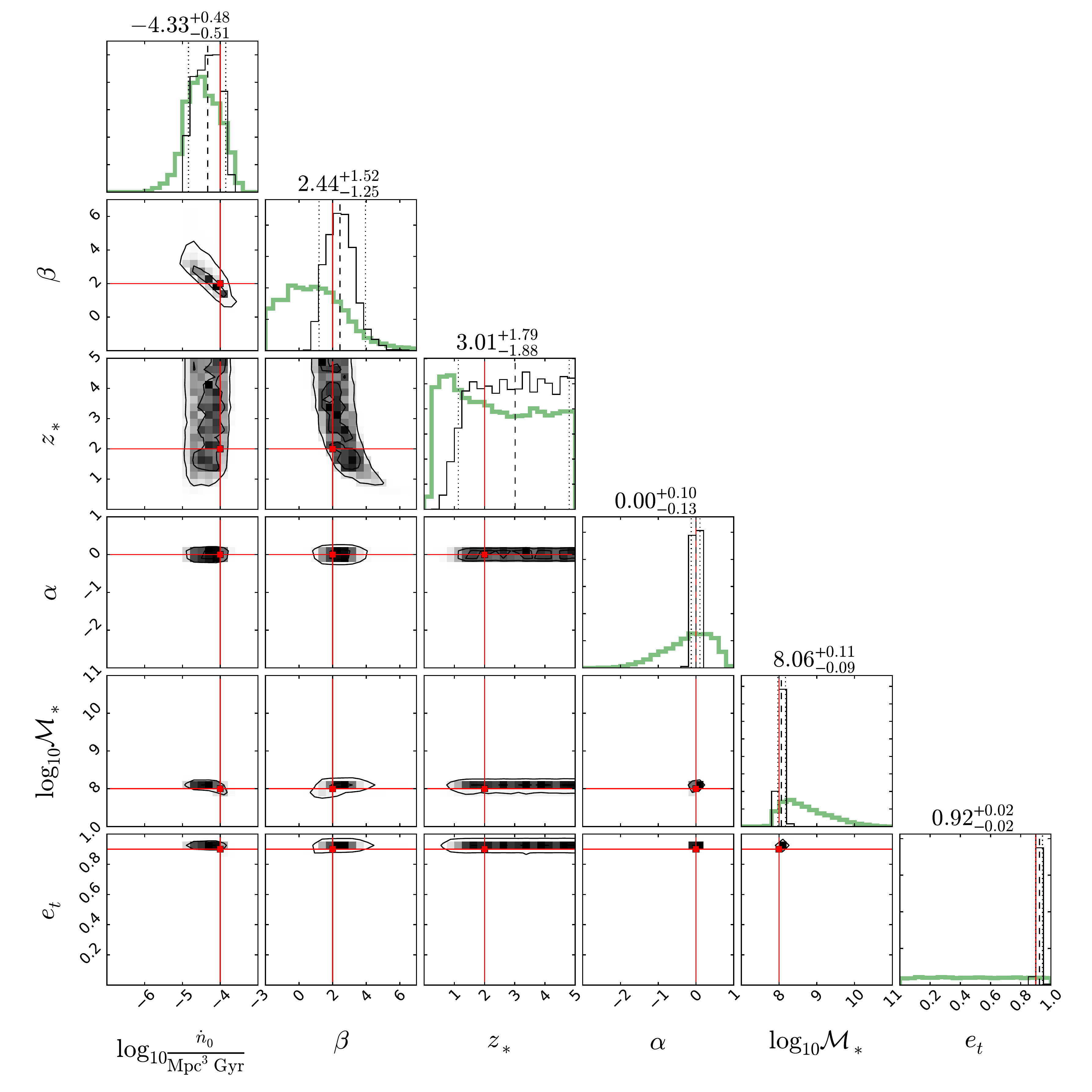}\end{subfigure} \\
\caption{Effects of imposing external constraints on the MBHB mass and redshift distribution on the science return of PTA observations. The injected model has default mass function parameters and $e_t=0.9$. In the left column we consider a moderate S/N detection with the {\it IPTA30} array, whereas the right panel is for an ideal detection as reported in Fig. \ref{fig:cornerplot_ideal}. Panel sequence and style are as in Fig. \ref{fig:detection_circ}. The additional dotted areas represent the restricted prior based on the astrophysical models of \protect\cite{Sesana:2013}. The thick green histograms in the bottom panels show the marginalised prior distribution on the model parameters once the astrophysical constraint is imposed (see main text for full details).
}
\label{fig:restricted_cases}
\end{figure*}

We saw in the previous section that parameter degeneracies prevent a precise characterization of the properties of the underlying MBHB population. This is because the GWB spectrum does not present sufficient structure to allow proper parameter estimation. In principle, the high frequency steepening of the GWB offers a tantalizing possibility of an independent measurement of the mass function parameters. In practice, unfortunately, the steepening generally occurs at $f>30$ nHz where PTA sensitivity drops significantly. A measurement might be possible for MBHB population featuring a heavy-biased MBHB mass function, for which the steepening occurs already around $f\approx 10$ nHz. However, even in this case, errorbars on the detected amplitude at the highest frequency bins would be quite large, making a proper measurement of the drop problematic.    

Although this is likely out of reach for current and planned PTA efforts, as a proof of principle we show here what information can be recovered with a measurement of the GWB spectrum up to $f=5\times10^{-7}$ Hz, possible with our {\it ideal} array. Performing a parameter space exploration would be impractical, because for 30 years of observation, the signal would be observed in about 500 frequency bins, making the evaluation of the likelihood function prohibitively time consuming. We therefore interpolate the observations (with relative errorbars) in 20 equally log-spaced bins in the range $10^{-9}-5\times10^{-7}$ Hz. Note that the total S/N of such detection is not much higher than the {\it SKA20}, however we will see that the high frequency extension makes a critical difference in the recovery of the MBHB population parameters (even if we are not using all the information enclosed in the original 500 frequency bins). This is shown in Fig. \ref{fig:cornerplot_ideal} for our standard MBHB population with $e_t=0.01$ (left column) and $e_t=0.9$ (right column). The upper panels show that, contrary to all previous cases, the high frequency steepening is now well characterized; this is the key element, because its shape depends on the MBHB mass function. Posterior distributions of the population parameters are shown in the lower triangle plots. The parameters defining the MBHB mass function are now well constrained and peak around the injected values; the cut-off mass scale $\Mc_*$ is determined within a factor of three and the slope $\alpha$ within $\approx$0.2. The recovery of the eccentricity is also much cleaner. Posteriors are still broad, but in the circular case one can confidently say that the typical eccentricity of the MBHBs is $<0.16$ (95\% confidence) although the posterior peaks at $e_t\approx0.1$. This is because a non detection of a low frequency turnover is still consistent with mildly eccentric binaries at decoupling, even if the mass function parameters are fairly well determined. Similarly, for the eccentric case, one can state with $95\%$ confidence that the typical eccentricity of the MBHBs is $>0.7$ and the posterior is quite flat in the range $0.75<e_t<0.95$. One last thing to notice is that, even though the GWB spectrum is pinned down essentially exactly, there remains a remarkable uncertainty in the determination of the overall merger rate density $\ndot$. This is because of its intrinsic (not shown) degeneracy with the $\beta$ and $z_*$ parameters defining the redshift distribution of mergers. A low $\ndot$ normalization with a steep, positive redshift dependence $\beta$ can result in the same GWB as a much higher  $\ndot$ normalization with a flatter redshift dependence. Unless external information (see below) about the redshift evolution of the merger rate density is available, this degeneracy is unlikely to be disentangled on the basis of GWB measurements alone.

\subsection{Adding independent constraints}

So far, we considered what astrophysical information can be extracted by PTA observation only, deliberately ignoring any constraints on the MBHB population imposed by other observations. The motivation behind this agnostic choice is that those constraints are inevitably indirect, and involve either the rate of merging {\it galaxies} \citep[e.g.][]{2004ApJ...617L...9L,2011ApJ...742..103L,2012ApJ...747...85X} or the determination of the mass function of {\it single MBHs} \citep[see for example][]{2004MNRAS.354.1020S}. The conversion of a galaxy merger rate into a MBHB merger rate implies a number of uncertain assumptions about the relation between galaxy hosts and MBHs \citep[][and references therein]{2013ARA&A..51..511K} and the effectiveness of the MBHB coalescence following galaxy mergers \citep[e.g.][]{2014ApJ...789..156M,2016MNRAS.tmp.1488K}; on the other hand, the mass function of individual MBHs in galaxy centers does not provide direct information on the properties of {\it merging} MBHBs.

It is nevertheless instructive and interesting to fold those indirect constraints into the analysis to understand to what extent PTA observation can improve the current state of the art of MBHB knowledge. \cite{Sesana:2013} constructed a compilation of observationally-based MBHB merger distributions encompassing a wide uncertainty range in the galaxy merger rate and galaxy host-MBH relations. The outcome of the procedure is a loosely constrained MBHB mass function and redshift distribution resulting in a predicted GWB spanning almost two order of magnitudes in amplitude (at 99.7\% confidence). In general, in the best constrained areas (chirp masses in the range $10^7\Msol$-$10^{8.5}\Msol$ and $z<1.5$), the uncertainty range spans about two orders of magnitudes. To incorporate this information in our analysis, we draw a large sample of populations from our unrestricted parameter range, and we accept only those for which the mass and redshift functions fall within the range constrained by the \cite{Sesana:2013} models to update our prior. The restricted MBHB mass and redshift functions resulting from this procedure are shown as dotted areas in the central panels of Fig. \ref{fig:restricted_cases}. The restricted marginalized priors on the model parameters are shown in the triangular plots in the bottom panels and their median values and 90\% confidence intervals are listed in table \ref{tab1}. Furthermore, since the merger rates do not constrain the MBHB eccentricity distribution at decoupling, we assume a flat prior on $e_t$. The resulting prior GWB spectrum is shown in the upper panels of Fig. \ref{fig:restricted_cases}. As expected the range of $h_c$ is consistent with what is shown in figure 2 of \cite{Sesana:2013}. The difference in shape is due to the inclusion of the high frequency drop, and to the fact that we allow for very eccentric MBHB population, that cause a widening of the allowed $h_c$ range at the low frequency. We assume that the true underlying MBHB population is described by our default models (shown with dashed lines), that falls well within the restricted prior range, and that MBHBs have $e_t=0.9$ at decoupling.

\begin{table}
\begin{center}
\def\arraystretch{1.5}
\begin{tabular}{c|ccc}
\hline
parameter & prior & {\it IPTA30} & {\it ideal}\\
\hline
log$_{10}\dot{n}$     & $-4.47^{+0.73}_{-0.70}$& $-4.35^{+0.71}_{-0.61}$& $-4.43^{+0.48}_{-0.51}$\\
$\beta$              & $ 0.81^{+3.29}_{-2.43}$& $ 0.75^{+3.12}_{-2.41}$& $ 2.44^{+1.52}_{-1.25}$\\
$z_*$                & $ 2.39^{+2.36}_{-1.95}$& $ 2.45^{+2.28}_{-2.00}$& $ 3.01^{+1.79}_{-1.88}$\\
$\alpha$             & $-0.11^{+0.75}_{-1.25}$& $-0.09^{+0.52}_{-0.89}$& $ 0.00^{+0.10}_{-0.13}$\\
log$_{10}{\cal M}_*$  & $ 8.58^{+1.25}_{-0.65}$& $ 8.18^{+0.64}_{-0.25}$& $ 8.06^{+0.11}_{-0.09}$\\
$e_t$                & $ 0.50^{+0.45}_{-0.45}$& $ 0.78^{+0.20}_{-0.72}$& $ 0.92^{+0.02}_{-0.02}$\\

\hline
\end{tabular}
\caption{List of model parameters credible intervals for our constrained models. Each entry reports the median value together with the errors bracketing the 90\% confidence regions. The three columns list the values defined by our restricted prior, and the posterior values as measured by the arrays {\it IPTA30} and {\it ideal}.}
\label{tab1}
\end{center}
\end{table}

The results of the analysis for two different PTAs are shown in Fig. \ref{fig:restricted_cases} and measured parameter values are also listed in table \ref{tab1}. PTA observations in the foreseeable future ({\it IPTA30} case, left column) will place significant constraints to the higher end of the mass function, reducing the uncertainty range by more than one order of magnitude at ${\cal M}>10^8\Msol$. The redshift function is poorly constrained, because the mass integral of the merger rate is dominated by the abundance of MBHBs with ${\cal M}<10^8\Msol$, which remains poorly determined. This is also confirmed by the marginalised posterior distributions in the model parameters shown in the bottom panel. The posteriors on the overall merger rate $\ndot$ and on the redshift parameters $\beta$ and $z_*$ are essentially unaltered when compared to the prior, conversely, the prior knowledge of $\Mstar$ is significantly updated with a 90\% confidence interval shrinking by an order of magnitude. Note that, since $\Mstar$ is decently constrained, the detection of the low frequency turnover is now quite informative, and eccentric binaries are favoured, with a posterior probability distribution correctly peaking around $e_t=0.9$. In the ideal case, shown in the right column, the mass function is constrained almost exactly, and also our knowledge of the redshift evolution of the merger rate is significantly updated. The posterior distributions of the model parameters show that $\alpha, \Mstar$ and $e_t$ are pinned down with high accuracy. Moreover, also the degeneracy between the rate normalization and the redshift evolution is partially broken. The 90\% credible interval on $\dot{n}_0$ shrinks by a factor of three compared to the prior, and the slope of redshift dependence $\beta$ can be fairly well constrained, with a posterior peaking close to the injected value. This latter measurement is particularly interesting, because it would allow a direct comparison to the galaxy merger rate that is often observationally parametrised as being proportional to $(1+z)^{\beta}$. 

\section{Conclusions and outlook}
\label{sec:Conclusions}

We have performed the first extended investigation of the inverse problem for PTA data analysis, namely: given a PTA observation or upper limit, what constraints can be placed on the astrophysical properties of the underlying MBHB population? Our work expands on M16, by considering future detections in a sizeable frequency range, allowing us to fold into the analysis the information carried by the observed spectral shape of the GWB. To do so, we employed the semi-analytical model of \cite{2016arXiv161200455C} that describes the MBHB population model with six physical parameters: five parameters shaping the redshift dependent mass function and an additional eccentricity parameter $e_t$ that encapsulates the main effect of the MBHB coupling with the stellar environment. Depending on those parameters, the resulting GWB spectrum might show a significant departure from the nominal $f^{-2/3}$ power-law both at high frequency, due to small number statistics of the systems contributing to the signal \citep{SesanaVecchioColacino:2008}, and at low frequency, because of high eccentricity caused by interaction with stars in the inspiral phase. We explored to what extent such spectral features are recognizable and can be exploited to extract information from PTA observations. We assumed uninformative prior ranges in all the model parameters, consistent with the current absence of any secure direct observation of sub-parsec MBHBs emitting in the PTA relevant range.

We first used our analysis framework to assess the impact of current PTA upper limits, recovering the results of M16. In essence, a non detection can only impose a upper-bound on the overall merger rate density of MBHBs. Current PTA limits set this upper-bound to $\ndot<2.5\times10^{-3}$ (95\% confidence), which is close to the range of currently measured galaxy merger rate densities, indicating that PTA observation are getting into the interesting astrophysical range. We then extended our investigation to a number of future detection scenarios: an IPTA-like array ({\it IPTA30}), and SKA-like array ({\it SKA20}) and an ideal array with 500 pulsars at sub-ns precision ({\it ideal}). In all cases, a GWB observation will provide a solid measurement of the overall merger rate densities of MBHBs, with other model parameters being constrained to different degrees depending on the array. We found a strong degeneracy between $e_t$ and the typical mass scale of merging MBHBs, defined by the parameter $\Mstar$. The degeneracy can be broken only with a confident detection of the high frequency drop of the spectrum, which depends on the underlying mass function but not on the eccentricity at decoupling. Unfortunately, this is possible only if the signal is detected at $f\gtrsim$few$\times 10^{-8}$Hz, which might be out of range even for the SKA. Finally, we considered the benefit of PTA detection when priors on the MBHB mass functions provided by independent observations are folded into the analysis. We found that, in this case, even in the {\it IPTA30} case, the eccentricity parameter can be constrained, because the constrained prior allows a better measurement of the typical MBHB mass scale. Therefore, when combined with independent observations, PTA observations in the foreseeable future have the potential of greatly enhancing our knowledge of MBHB astrophysics and dynamics.

These results are subject to a number of caveats, that will be explored in future work. First, we did not considered measurement errors in the observations. Although we included uncertainties in the measured characteristic amplitude at each frequency, we centred them at the value of the injected signal. Including an additional scatter, will make the reconstruction of the spectrum more cumbersome, especially in the case of low S/N detection. Second, we did not include the intrinsic scatter of the signal amplitude due to the stochastic nature of the GWB. In our model, each set of parameters produces a single $h_c(f)$. However, the exact value of the GWB at each frequency depends on the statistics of rare massive systems, and therefore, each set of underlying MBHB population parameters produced a probability distribution of $h_c(f)$ at each frequency. This can be taken into account with a suitable modification of the likelihood function that we plan to implement as next step of this investigation. Finally, our current analysis is limited to the stochastic part of the signal. Especially at high frequency, bright sources will be individually resolvable, carrying a great deal of information about the most massive systems that can be used to complement the information provided by the GWB spectral shape. All these shortcomings can be addressed within our framework via suitable modifications of different stages of the pipeline, and will be the subject of future publications in this series of papers.

\section*{acknowledgements}
We acknowledge the support of our colleagues in the European Pulsar Timing Array. A.S. is supported by a University Research Fellow of the Royal Society.

\bibliographystyle{mnras}
\bibliography{bibliography}

\label{lastpage}

\end{document}